%% file: NGC3115.tex
\documentclass[fleqn,usenatbib]{mnras}

\usepackage{newtxtext,newtxmath}

\usepackage[T1]{fontenc}
\usepackage{tablefootnote}
\usepackage{ae,aecompl}
\usepackage{multirow}
\usepackage{amsmath}
\usepackage{booktabs}
\usepackage{arydshln}
\usepackage[normalem]{ulem}
\usepackage{threeparttable}
\newcommand{\ra}[1]{\renewcommand{\arraystretch}{#1}}
\usepackage{graphicx}	
\usepackage{amsmath}	
\usepackage{soul}
\usepackage{color}
\usepackage{soul}

\usepackage{newtxtext,newtxmath}

\title[The formation history of NGC 3115]{Recovering the origins of the lenticular galaxy NGC 3115 using multi-band imaging}

\author[M. L. Buzzo, A. Cortesi et al.]{Maria Luisa Buzzo$^{1}$\thanks{E-mail: maria.buzzo@usp.br},
Arianna Cortesi$^{2,1}$,
Jose A. {Hernandez-Jimenez}$^{3,1}$,
Lodovico Coccato$^{4}$,
Ariel Werle$^{5,1}$,
\newauthor
Leandro {Beraldo e Silva}$^{6}$,
Marco Grossi$^{2}$,
Marina Vika$^{7}$,
Carlos Eduardo Barbosa$^{1}$,
Geferson Lucatelli$^{8}$,
\newauthor
Luidhy {Santana-Silva}$^{9}$,
Steven Bamford$^{10}$,
Victor P. Debattista$^{6}$,
Duncan A. Forbes$^{11}$,
Roderik Overzier$^{12,1}$,
\newauthor
Aaron J. Romanowsky$^{13,14}$,
Fabricio Ferrari$^{8}$,
Jean P. Brodie$^{11,13}$,
Claudia {Mendes de Oliveira}$^{1}$
\\ \\
$^{1}$ Universidade de S\~ao Paulo, IAG, Rua do Mat\~ao 1226, Cidade Universit\'aria, S\~ao Paulo 05508-900,Brazil\\
$^{2}$ Observat\'orio do Valongo, Ladeira do Pedro Ant\^onio 43, CEP:20080-090, Rio de Janeiro, RJ, Brazil\\
$^{3}$ Departamento de Ciencias F\'isicas, Universidad Andres Bello, Fernandez Concha 700, Las Condes, Santiago, Chile \\
$^{4}$ European Southern Observatory, Karl-Schwarzchild-str., 2, 85748 Garching b. Munchen, Germany \\
$^{5}$ INAF - Osservatorio Astronomico di Padova, Vicolo dell'Osservatorio 5, 35122 Padova, Italy  \\
$^{6}$ Jeremiah Horrocks Institute, University of Central Lancashire, Preston, PR1 2HE, UK \\
$^{7}$ Institute for Astronomy, Astrophysics, Space Applications \& Remote Sensing, National Observatory of Athens, Penteli, 15236, Athens, Greece \\
$^{8}$ Instituto de Matematica Estatistica e Fisica, Universidade Federal do Rio Grande (IMEF-FURG), Rio Grande, RS, Brazil \\
$^{9}$ NAT-Universidade Cruzeiro do Sul / Universidade Cidade de S\~ao Paulo, Rua Galv\~ao Bueno, 868, 01506-000, S\~ao Paulo, SP, Brazil \\
$^{10}$ School of Physics \& Astronomy, The University of Nottingham, University Park, Nottingham, NG7 2RD, UK \\
$^{11}$ Centre for Astrophysics \& Supercomputing, Swinburne University, Hawthorn VIC 3122, Australia \\
$^{12}$ Observatorio Nacional, Rua Jos\'e Cristino, 77. CEP 20921-400, S\~ao Crist\'ov\~ao, Rio de Janeiro-RJ, Brazil\\
$^{13}$ Department of Physics \& Astronomy, San Jos\'e State University, One Washington Square, San Jose, CA 95192, USA \\
$^{14}$ University of California Observatories, 1156 High St., Santa Cruz, CA 95064, USA 
}

\date{Accepted 2021 March 30. Received 2021 March 30; in original form 2020 December 31.}

\pubyear{2020}

\begin{document}
\label{firstpage}
\pagerange{\pageref{firstpage}--\pageref{lastpage}}

\maketitle

\begin{abstract}
A detailed study of the morphology of lenticular galaxies is an important way to understand how this type of galaxy formed and evolves over time. Decomposing a galaxy into its components (disc, bulge, bar, ...) allows recovering the colour gradients present in each system, its star formation history, and its assembly history. We use {\sc GALFITM} to perform a multi-wavelength structural decomposition of the closest lenticular galaxy, NGC 3115, resulting in the description of its stellar light into several main components: a bulge, a thin disc, a thick disc and also evidence of a bar. We report the finding of central bluer stellar populations in the bulge, as compared to the colour of the galaxy outskirts, indicating either the presence of an Active Galactic Nucleus (AGN) and/or recent star formation activity. From the spectral energy distribution results, we show that the galaxy has a low luminosity AGN component, but even excluding the effect of the nuclear activity, the bulge is still bluer than the outer-regions of the galaxy,  revealing a recent episode of star formation. Based on all of the derived properties, we propose a scenario for the formation of NGC 3115 consisting of an initial gas-rich merger, followed by accretions and feedback that quench the galaxy, until a recent encounter with the companion KK084 that reignited the star formation in the bulge, provoked a core displacement in NGC 3115 and generated spiral-like features. This result is consistent with the two-phase formation scenario, proposed in previous studies of this galaxy.
\end{abstract}


\section{Introduction}

Lenticular galaxies, also known as S0s, were first introduced as an intermediate class of galaxies in the Hubble classification scheme \citep{Hubble1936}, and have been since studied in a large range of environments, redshifts and masses, in order to understand the processes involved in morphological transformations. Consequently, it has been long established their close relation to high density environments \citep{VanGorkom,Mo}.
In such environments, the number of lenticular galaxies increases toward the center of the cluster, in an opposite trend to the number of spiral galaxies \citep{Dressler80}. However, lenticular galaxies are found in all environments, corresponding to a fraction of 8\% in the field, 13\% in groups and 28\% in the dense Virgo cluster \citep{VandenBergh}.

The number of lenticular galaxies varies also with redshift, i.e. S0s are more common at lower redshifts, while the number of spiral and star forming galaxies grows with increasing redshift \citep{Dressler97, Barro}. 
Lenticular galaxies, moreover, account for the larger number of high-mass galaxies in the local universe \citep{Bernardi}. However, the processes that lead to the formation of lenticular galaxies are still an open field of research.

Different formation scenarios would leave different imprints on the galaxy kinematics and stellar populations. In lenticular galaxies that are the result of spirals undergoing ram-pressure stripping \citep{Gunn&Gott}, we might expect the disc stellar population to be older than the one of the spheroid \citep{Johnston}, either because the gas truncation leads to an outside-in quenching of the star formation \citep{Schaefer, Finn}, or because the bulge stellar population could be rejuvenated by a last star formation episode \citep{BekkiCouch}, caused by the collapse of the stripped material into the central region. Lenticular galaxies that are formed through the merger and/or accretion of companion galaxies and satellites, might have younger discs than the bulge due to the accretion of gas rich satellites onto the disc, which will enhance the star formation \citep{Diaz}. Finally, in lenticulars that are relics of high-redshift galaxies, such as primordial galaxies, which never passed through a spiral phase, the bulge and the disc would have the same stellar population, as they are formed by the same material \citep{SahaCortesi}. 

The mechanisms that form lenticular galaxies in low density environments would not be the same as the ones creating their high-density counterparts, since mergers are more likely to occur in low density environments \citep{Eliche-Moral,Tapia}. Tidal interactions are also typical of galaxy groups, while hot gas and a deep gravitational well are necessary to harass or strip a galaxy infalling a cluster of galaxies. Cluster lenticular galaxies might have been pre-processed in group or field environments, and it is therefore harder to isolate the mechanism responsible for their formation. On the contrary, isolated lenticular galaxies provide an ideal laboratory for studying the formation of this intriguing class of galaxies. 

Using integral field unit (IFU) spectroscopy of a sample of 279 lenticular galaxies in the Mapping nearby galaxies at apache point observatory (MaNGA) survey, \cite{McKelvie2018}, have shown that lenticular galaxies have different stellar populations depending on their masses, such that low mass galaxies ($M_{\star} < 10^{10} M_{\odot}$) are probably quenched spiral galaxies, whereas high mass lenticulars  ($M_{\star} > 10^{10} M_{\odot}$) are the result of merger events, either minor \citep{Bekki,Bournaud} or major \citep{Eliche-Moral, Querejeta, Tapia, Borlaff}, independently of the environment. However, \cite{Coccato} showed that the kinematics of lenticular galaxies living in low and high density environments are different at $1.5 \sigma$  confidence level at $1 R_{e}$, suggesting that both mass and environment play a role in shaping the fate of the progenitor(s) of lenticular galaxies. It is still unclear, however, the exact role of these two factors in the formation of S0 galaxies.

In this work, we aim to answer the question of how the closest field lenticular galaxy, NGC 3115, was formed. This galaxy has been studied in a wealth of projects using different data sets \citep{Dolfi, Guerou, Poci, 3115_3}. NGC 3115 is the nearest galaxy hosting a billion solar mass black hole \citep{Kormendy96, Emsellem} and several works have shown that it has a weak AGN activity \citep{Lauer, Wrobel}. Moreover, there is a general agreement that NGC 3115 formed through a two phase formation scenario, and it is quenched and isolated \citep{Arnold11, Guerou, Poci}. Yet, several aspects of its formation are still unsolved, such as the presence of a low-luminosity AGN \citep{Almeida}, as well as of a large sample of UCDs and companion galaxies \citep{Dolfi}. Did they play a role in the quenching mechanism and how did they affect the galaxy kinematics and structure? For example, by recovering the galaxy kinematics using Planetary Nebulae, \citet{Cortesib} showed that the disc of NGC 3115 is rotationally supported, but that the value of V$/\sigma$ is lower than what found for spiral galaxies, suggesting that this galaxy is not simply a faded spiral. Similar results are found by studying the globular cluster population \citep{Zanatta} and integrated stellar light \citep{Arnold11}. 

Spectroscopic data allows for precise measurements of age and metallicity of the stars in a galaxy, but can generally only cover a small wavelength range. Spectroscopy is also limited in aperture, 
even IFU observations are usually restricted to the central parts of the galaxies, within a few effective radii.
Furthermore, the large observing times required for IFU spectroscopy restrict this technique to small samples, whereas multi-band observations may be more easily used to provide volume-limited, representative samples.
A compromise can be achieved with narrow-band photometric surveys \citep[e.g][]{SPLUS,JPLUS,JPAS}, which provide large number statistics and allow us to break some of the degeneracies involved in recovering the formation history of galaxies.

Spectral energy distributions (SEDs) obtained from broad-band photometry are far less restricted than IFU observations in spatial coverage, and the ability of this technique to gather information from different wavelengths, from ultraviolet to infrared, make up for the loss of detailed $\lambda$-by-$\lambda$ constraints provided by spectroscopic data \citep{Salim2007, MAGPHYS, Prospector,Werle2019}.

We provide in this work a study of a mimicked multi-wavelength image, followed by a structural decomposition of NGC 3115. Previous works \citep{VIKA13, VULCANI, Dimauro18, Dimauro19,Psychogyios,Mosenkov} have demonstrated the ability to decompose galaxies into their different structural components consistently and accurately, indicating that such decomposition may provide a more detailed picture of the formation history of the galaxy than by considering a galaxy as a whole.
This pilot study of NGC 3115 was achieved by compiling ultraviolet, optical, near infrared, and infrared data covering the galaxy out to 6.5 arcminutes from its centre, thus studying the system out to $\sim7$ effective radii \citep[Re = 58.1 arcsec][]{Cortesib}, which provided a large scale perspective of its main structural components. We then performed a detailed, consistent, multi-wavelength decomposition of the main structural components of the galaxy. 
With that, we expect to enhance the understanding of the galaxy formation and dynamical history and the role of the environment on its evolution.

The paper is structured as follows: in \textsection \ref{sec:data} we describe the data gathered for this study, as well as the process of the registration, calibration and PSF homogenisation. In \textsection \ref{sec:methods} we explain the {\sc GALFITM} setup and the details of the models used for the decomposition. In \textsection \ref{sec:results_galfitm} we show all results obtained with {\sc GALFITM}. In \textsection \ref{sec:analysis} we report our analysis. In \textsection \ref{sec:discussion} we discuss the results and propose a formation history scenario for NGC 3115. 

Throughout this work, we adopt the distance to NGC 3115 of D $= 9.4$ Mpc \citep{Brodie14}. We assume a standard $\Lambda$CDM model, with H$_0 = 70.5$ km s$^{-1}$ Mpc$^{-1}$ \citep{Komatsu}. 

\section{Data}
\label{sec:data}

\subsection{Data sources and registration}
To study NGC 3115, we use 11 images of the galaxy obtained with different telescopes, with the intent of covering a wide wavelength range, not provided by any survey currently. We collect data from the Galaxy Evolution Explorer \citep[GALEX]{GALEX}, Subaru Suprime Cam \citep{SUBARU}, Dark Energy Cam \citep[DECam]{DECAM}, 2MASS \citep{2MASS} and Wide-Field Infrared Survey Explorer \citep[WISE]{WISE}. In Table \ref{table:data} we show the details of each image used in this work.

\begin{table}
\caption{Details about the archival data used in this work}
\scalebox{0.9}{
\begin{tabular}{lcccc}
\hline
 \textbf{Instrument} &  \textbf{Band} & \textbf{$\lambda$ (\AA)} & \textbf{Pixel scale ('')} & \textbf{Zero point} \\ \hline \hline
 GALEX & FUV & 1520 & 1.50 & 18.82\\
 GALEX & NUV & 2270 & 1.50 & 20.08\\
 Subaru Suprime Cam & $g$ & 4770 & 0.20 & 30.47\\
 Subaru Suprime Cam & $r$ & 6800 & 0.20 & 30.50\\
 Subaru Suprime Cam & $i$ & 7630 & 0.20 & 31.50\\
 DECam & $z$ & 9260 & 0.26 & 27.80\\
 2MASS & J & 12500 & 1.00 & 20.81 \\
 2MASS & H & 16500 & 1.00 & 21.88\\
 2MASS & Ks & 21700 & 1.00 & 21.87\\
 WISE & 3.4 & 34000 & 2.75 & 23.20\\
 WISE & 4.6 & 46000 & 2.75 & 22.84\\ \hline
\end{tabular}}
\label{table:data}
\end{table}

To create a consistent multi-band image with images obtained with different instruments, we did a thorough treatment of the data before starting the fit. The images were all set to have the same pixel scale, world coordinate system (WCS), dimension and centre as the 2MASS H band image (1''/px). This was done using {\sc Montage} \citep{MONTAGE}, a tool for gathering fits images into custom mosaics using a reference image, in our case, the H band of 2MASS. 
We choose to use the H band as a reference, because the pixel scale is 1 arcsec, which is easier to deal with, and also this pixel scale is a compromise point between the small pixel scale of Subaru (0.2 arcsec) and large of WISE (2.75 arcsec).
Differences in the gain, point spread function (PSF) and seeing are taken into account in the process of fitting with {\sc GALFITM}. 

We created a synthetic PSF of each image using the Moffat distribution \citep{Moffat}, which is based on a softened exponential profile to model the PSF. The parameters used to create such PSFs were obtained using the {\sc imexamine} function in {\sc IRAF}.

To measure the sky background, we used the median value of small regions in the images, with little contamination from stars, followed by a sigma clipping, until convergence to the best value was reached \citep{Jose1,Jose2}.
The best sky value for each wavelength band is defined and fixed in the configuration file of {\sc GALFITM} to be considered by the routine when creating the best fit model.

The zero points of these images were retrieved from each respective telescope, all in AB magnitudes. However, for the Subaru Suprime Cam images, these values were not available.
To retrieve such information, we used the detailed study of the globular cluster (GC) system around NGC 3115, which made use of the same images analysed in this work, presented in the paper of \cite{GCs}. Comparing the magnitudes of each object in the paper and our determinations of the instrumental magnitudes, we have estimated the offset, which corresponds to the zero points of these images. These are $31.40 \pm 0.28$ mag, $30.5 \pm 0.28$ mag, $31.5 \pm 0.76$ mag for \textit{g}, \textit{r} and \textit{i}, respectively. We also recovered the zero points by using DECam images in the \textit{g}, \textit{r} and \textit{i} bands as a reference and we obtained the same values as from the GCs. 

\section{Method: setting of {\sc GALFITM}}
\label{sec:methods}
Historically, images of galaxies have been modelled in each passband individually, or aperture photometry was used to extract galaxy's luminosity into an area defined by the same radius in each band. This type of analysis might lead to inconsistent colours and photometric properties due to different image resolution and depth \citep[see][]{VIKA14}.
To overcome such a problem, {\sc Megamorph} \citep{Haussler2013,VIKA13} was developed to perform simultaneous multi-wavelength fitting of images.
In particular, this technique allows to simultaneously fit bands that have lower signal-to-noise ratios (S/N), like ultraviolet and infrared data, with images with high S/N, as optical deep images, resulting in consistent and better-constrained parameters. This is obtained by expanding each fitted parameter in a wavelength-dependent series, using a smooth polynomial function (Chebyshev polynomial), where the maximum order of the series (hereafter orders) must be at most equal to the number of passbands. This method makes the fit possible and robust, even for low S/N bands.
{\sc Megamorph} is a project that has produced a new version of  {\sc GALFIT} \citep{GALFIT} and {\sc GALAPAGOS} \citep{GALAPAGOS} routines, respectively {\sc GALFITM} and  {\sc GALAPAGOS-2}. In this work, we make use of {\sc GALFITM} to perform the analysis.

\subsection{Tuning the initial configurations for {\sc GALFITM}}
\label{section:dof}

The fundamental steps in recovering the galaxy components using {\sc GALFITM} are: identifying the number of physically motivated components, identifying the order of the series associated with each parameter for every component and defining the initial conditions for every component. On one hand, the highest the number of free parameters (and the number of components) and the highest the number of orders, the lowest will be the reduced $\chi^2$ of the fit. On the other hand, the recovered components might be purely mathematical artefacts. Moreover, when including more than one component in the fit, we can assume that the colour variation inside each component is negligible when compared with the colour variation between different components. This approximation can be used to extract robust parameters from faint images, by anchoring the fit to the neighbouring deeper images. Again, to find a balance between freedom (i.e. large number of degrees of freedom) and no freedom at all (fixed aperture) is a fine-tuning process that requires trial and error. In the next subsections, we describe the two methods we developed to find the best number of fitted components, best orders combination and initial conditions of each component. 

\subsubsection{A frequentist approach: selecting the best fit model among 32000 realisations}
One important aspect in the {\sc GALFITM} modelling is the order of the series allowed in the fitting process, which dictates how much each parameter may vary as a function of the wavelength.
The higher the order of the series, the more flexible the function becomes, such that when the order is the same as the number of input images, that parameter can vary independently between bands as if the fit would be performed on every image separately. On the other hand, if the order of the series is 0, the best model would be exactly the one defined by the initial conditions in every band. 
Thus, to maximise the fit performance, we need to find the best combination of orders of each parameter.

\begin{figure*}
    \centering
    \includegraphics[width=\textwidth,trim=0 0 8cm 0,clip]{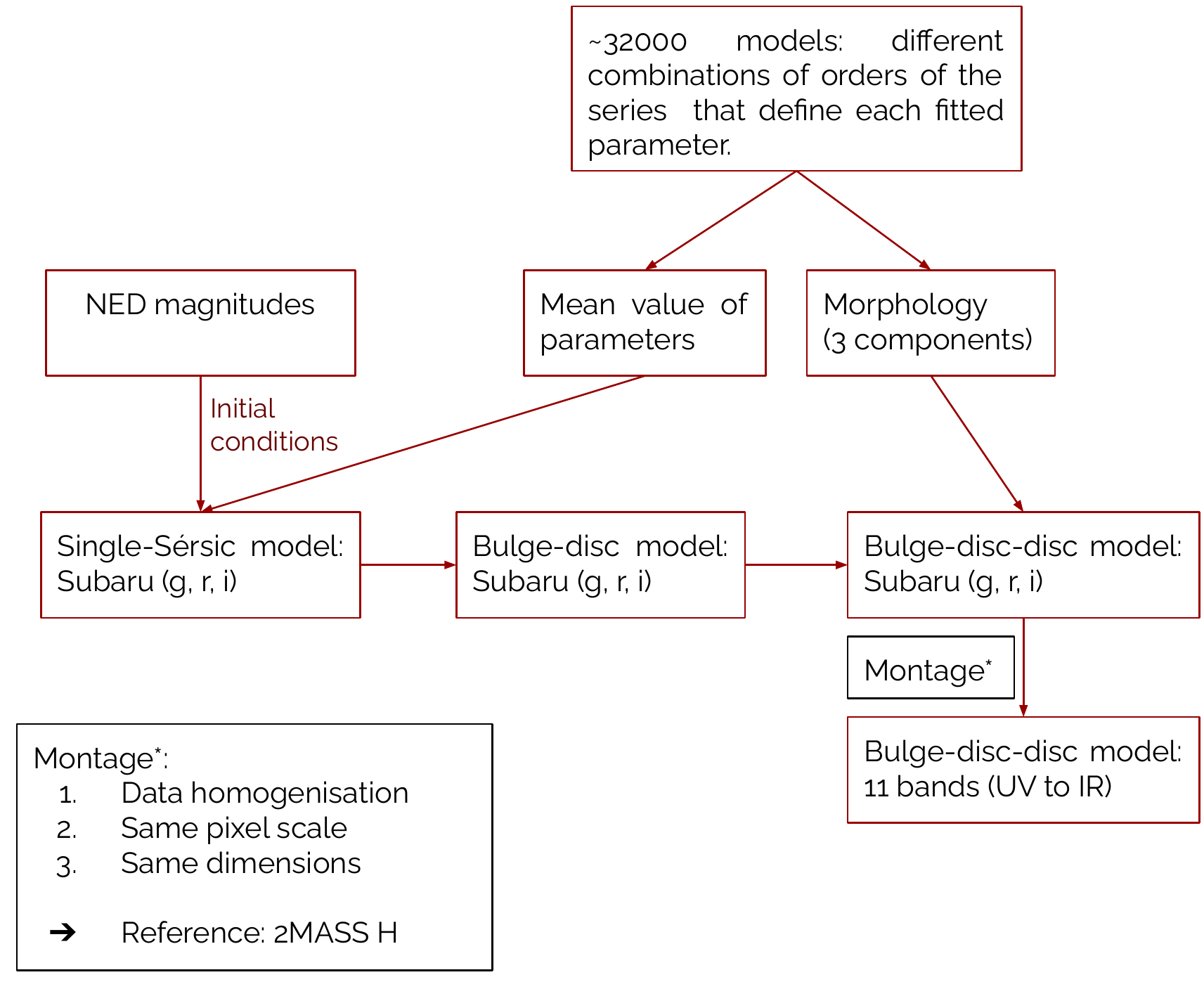}
    \caption{Steps taken for the creation of the best fit model of NGC 3115 with {\sc GALFITM}, using three components (one bulge and two discs) in 11 bands, from GALEX to WISE.}
    \label{fig:model_montage}
\end{figure*}

To do so, we created a set of $\sim$ 32 thousand {\sc GALFITM} models using different combinations of orders that define the variation of each parameter with wavelength, and an increasing number of fitted components (from a Single S\'ersic (SS) model to a bulge-disc (BD) decomposition, to a bulge and two discs (BDD) decomposition). The analysis of these 32 thousand {\sc GALFITM} models and the results obtained are shown in the Appendix \ref{sec:appendix_galfitm}.

Throughout the process of finding the best model for NGC 3115, we inspected not only the best combination of orders for the parameters but also the best set of components that would better capture the galaxy's morphology. We investigated from the most extreme case of three S\'ersic profiles until the other extreme of three exponential discs. The best fit model was the result of fitting a S\'ersic profile and two exponential disc profiles.
(A careful analysis of the results of the simulations of different combinations of orders is also shown in Appendix \ref{sec:appendix_galfitm}.)

By analysing the residual images and the BIC and AIC parameters we obtained the following results:
\begin{itemize}
    \item 3 components are needed to recover the morphology of NGC 3115 using this dataset.
    \item The best fit is obtained using low orders for each parameter.
    \item The best initial conditions are recovered.
\end{itemize}

Yet, the large number of realisations of the fit makes unfeasible a human evaluation of the residuals, which remains the most reliable source of judgement of the goodness of this type of fits.
Therefore, we refine the fitting procedure as explained in the following section. We keep in mind, though, that the best fit model was obtained by including three components (one S\'ersic profile and two exponential disks), with low orders of freedom of the fitted parameters and well defined initial conditions.

\subsubsection{An intuitive approach: using physical assumption and visual inspection to find the best fit model}

\begin{figure*}
    \centering
     \includegraphics[width=\textwidth]{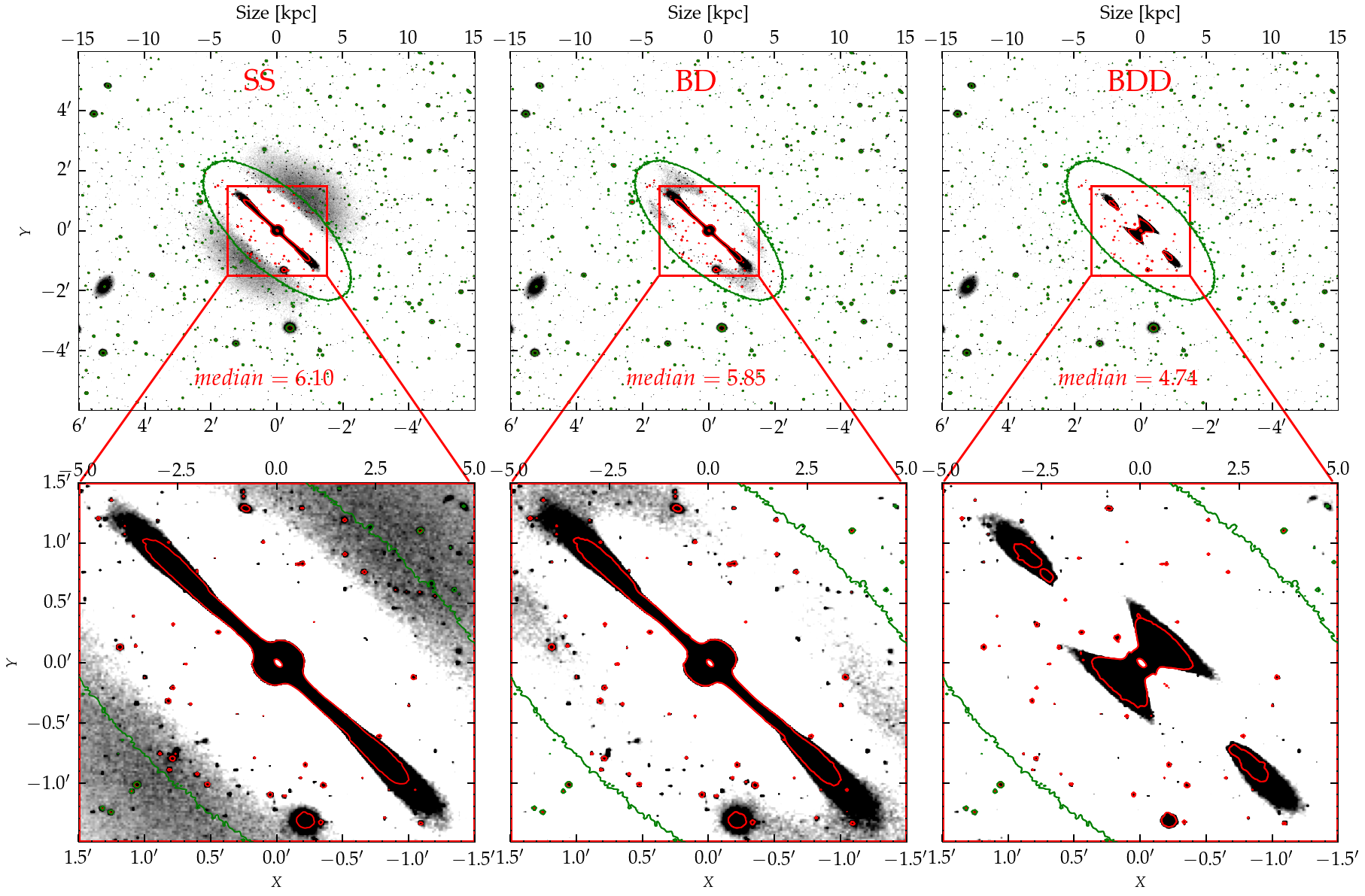}
     \caption{Evolution of the residuals (image - model) of the fit when using only one component (single S\'ersic (SS)), two components (bulge-disc (BD)) and three components (bulge-disc-disc (BDD)), from left to right. The top line shows the r-band image, while the bottom line shows the zoomed-in rectangles highlighted on the full images. The green contours show the outermost isophote of the modelled galaxy. The red contours highlight the position of the thin disc and the subsequent fit of this thin disc in the final panel. The median of the residuals of each fit is shown in the top panel in units of counts.}
     \label{evolution_residuals}
 \end{figure*}

An alternative approach, rather than generating a high number of models to statistically encounter the best one, is to make physically based assumptions to define the fitting parameters. For example, when including more than one component in the fit, we can assume that the colour variation inside each component is negligible when compared with the colour variation between different components. Such assumption translates into using a low number of orders for the fitted parameters, as also occurring in the case of the best fit model obtained from the simulations. Moreover, the deepest and with highest S/N images, the $g$, $r$ and $i$ images (Subaru), can be used to fix the shape of the galaxy components' at all wavelengths. 
The fit of the Subaru images started with a single S\'ersic model, evolving to a bulge-disc decomposition and finally to the bulge and two discs model - as found as the best model in the previous section.  
For the single S\'ersic model, we retrieved the initial magnitudes from the Nasa Extragalactic Database (NED) and the other initial parameters were set by the best single S\'ersic model out of the 32 thousand. Then, we used the outcome of the single S\'ersic model to set the initial conditions of the bulge-to-disc light decomposition and the results of the latter to define the initial conditions for the bulge and two discs model. 
Fig. \ref{evolution_residuals} shows the evolution of the residuals for the three fits (SS, BD, BDD). The best fit model is obtained by fitting three components, again confirming the results obtained in the previous section, by studying the residuals of 32000 models. 
We then adopt the best-fit model obtained for the three deep images of Subaru to fix the shape of the galaxy's components at all wavelengths, returning to use the images homogenised with {\sc Montage}. In other words, we use the outcome of the fit of the Subaru images to define the initial conditions for the fit of the 11 bands, allowing the parameters to vary with 1 order (i.e. they must converge to the same value in all bands), except for the magnitude and disc-scale length, which we allow to vary with orders 9 and 2 to define the series, respectively, for all components. In fact, an order of 9 for the magnitudes is large enough to ensure an independent evaluation of the parameter, but still low enough to decrease the error bars and recover a smooth profile. While the choice of 2 for the order of the disc-scale length was made to improve the residuals.

The parameters found for this best model, as well as the allowed order of each parameter are summarised in Table \ref{table_results_model}, in the results section (\textsection \ref{sec:results_galfitm}).
A summary of the steps taken to create the model that best describes NGC 3115 in all 11 bands is shown in Fig. \ref{fig:model_montage}.

\section{{\sc GALFITM} models}
\label{sec:results_galfitm}

\subsection{A bulge, thick disc and thin disc representation}

As explained in the previous section, to find the best model, we passed from a single S\'ersic model, to a bulge-disc model, to finally a bulge and two discs model, which provided the smallest residuals, while maintaining the physical reliability of the model, i.e. with physically meaningful components. 

\begin{figure*}
\includegraphics[width=\textwidth]{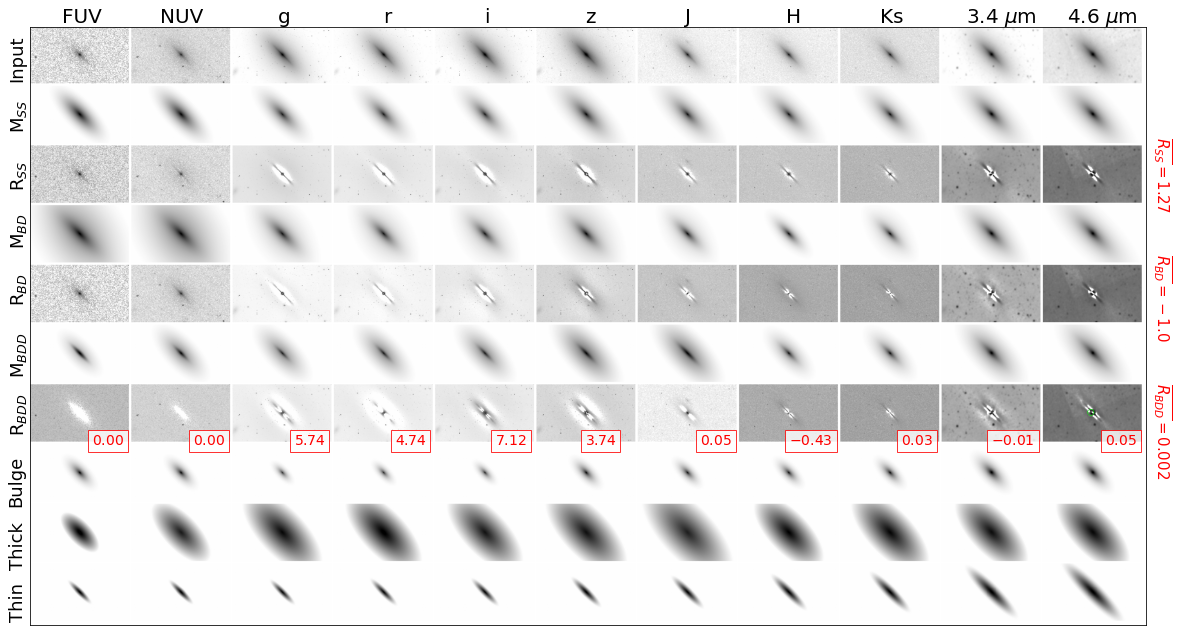}
\caption{{\sc GALFITM} models of NGC3115. First row: input images. Second and third rows: model and residual (data-model) of the model with one component (single S\'ersic (SS)). Fourth and fifth rows: model and residual (data-model) of the model with two components (bulge-disc (BD)). Sixth and seventh rows: model and residual (data-model) of the model with three component (bulge-disc-disc (BDD)). Annotated in red in the seventh row are the median value of the residuals for each band in the BDD model. Eighth, nineth and tenth rows correspond to the bulge, thick disc and thin disc models present in the BDD model, respectively. The columns represent each wavelength band. Annotated in red in the right-hand side of the image are the median residuals over all bands for the SS, BD and BDD models.}
\label{model_BDD}
\end{figure*}

\begin{table*}
\caption{Results from best fit model of NGC 3115. The table shows the parameters that define the best fit of each component of the galaxy: bulge, outer spheroid and thin disc. The order of the series of each parameter are shown in the third line of the table and describe how each parameter can vary with wavelength. See section \textsection \ref{section:dof}}
\hspace*{-0.5cm}
\scalebox{0.68}{%
\begin{tabular}{cccccccccccccccc}
\hline                                                                 
     & \multicolumn{5}{c}{Bulge} & \multicolumn{5}{c}{Thick Disc} & \multicolumn{5}{c}{Thin disc} \\ \hline
    & mag (AB)  & n  & Re ('') & b/a & PA & mag (AB)   & n & Rs ('')  & b/a  & PA  & mag (AB)   & n & Rs ('')  & b/a  & PA  \\ \hline
Orders & $9$    & $1$  & $1$  & $1$   & $1$  & $9$     & $0$   & $2$   & $1$    & $1$   & $9$     & $0$  & $2$   & $1$    & $1$   \\ \hline \hline
FUV &  $15.07 \pm 1.49$    &  $3.50 \pm 0.01$  & $21.68 \pm 0.02$   & $0.44 \pm 0.01$     &  $44.19 \pm 0.01$  &  $15.55 \pm 0.51$     &  $1$   &  $61.17 \pm 0.10$   & $0.45 \pm 0.01$      &  $43.42 \pm 0.02$    &  $15.18 \pm 0.41$     &  $1$  &   $23.13 \pm 0.02$  &  $0.22 \pm 0.01$    &  $44.83 \pm 0.02$   \\
NUV &  $15.73 \pm 0.72$   &  $3.50 \pm 0.01$  & $21.68 \pm 0.02$   & $0.44 \pm 0.01$     &  $44.19 \pm 0.01$  &  $15.25 \pm 0.72$     &  $1$   &  $61.60 \pm 0.12$   & $0.45 \pm 0.01$      &  $43.42 \pm 0.02$    &   $15.45 \pm 0.60$    &  $1$  &  $23.50 \pm 0.02$   &  $0.22 \pm 0.01$    &  $44.83 \pm 0.02$   \\
g   &  $12.33 \pm 0.02$  &  $3.50 \pm 0.01$  & $21.68 \pm 0.02$   & $0.44 \pm 0.01$     &  $44.19 \pm 0.01$  &    $10.56 \pm 0.02$   &  $1$   &  $63.05 \pm 0.09$   & $0.45 \pm 0.01$      &  $43.42 \pm 0.02$    &  $11.47 \pm 0.01$     &  $1$  &  $24.74 \pm 0.02$   &  $0.22 \pm 0.01$    &  $44.83 \pm 0.02$   \\
r   & $11.38 \pm 0.02$   &  $3.50 \pm 0.01$  & $21.68 \pm 0.02$   & $0.44 \pm 0.01$     &  $44.19 \pm 0.01$  & $9.61 \pm 0.01$      &  $1$   & $64.22 \pm 0.08$    & $0.45 \pm 0.01$      &  $43.42 \pm 0.02$    &  $10.36 \pm 0.01$     &  $1$  &   $25.75 \pm 0.02$  &  $0.22 \pm 0.01$    &  $44.83 \pm 0.02$   \\
i   & $11.30 \pm 0.03$    &  $3.50 \pm 0.01$  & $21.68 \pm 0.02$   & $0.44 \pm 0.01$     &  $44.19 \pm 0.01$  & $9.56 \pm 0.01$      &  $1$   &  $64.70 \pm 0.08$   & $0.45 \pm 0.01$      &  $43.42 \pm 0.02$    &   $10.30 \pm 0.02$    &  $1$  &   $26.16 \pm 0.02$  &  $0.22 \pm 0.01$    &  $44.83 \pm 0.02$   \\
z   &  $11.35 \pm 0.02$ &  $3.50 \pm 0.01$  & $21.68 \pm 0.02$   & $0.44 \pm 0.01$     &  $44.19 \pm 0.01$  &    $9.30 \pm 0.04$   &  $1$   &  $65.64 \pm 0.08$   & $0.45 \pm 0.01$      &  $43.42 \pm 0.02$    &  $9.97 \pm 0.02$     &  $1$  &  $26.96 \pm 0.02$   &  $0.22 \pm 0.01$    &  $44.83 \pm 0.02$   \\
J   &  $10.74 \pm 0.14$  &  $3.50 \pm 0.01$  & $21.68 \pm 0.02$   & $0.44 \pm 0.01$     &  $44.19 \pm 0.01$  &   $7.45 \pm 0.01$    &  $1$   &  $67.51 \pm 0.06$   & $0.45 \pm 0.01$      &  $43.42 \pm 0.02$    &   $7.56 \pm 0.01$    &  $1$  &   $28.57 \pm 0.02$  &  $0.22 \pm 0.01$    &  $44.83 \pm 0.02$   \\
H   &  $8.14 \pm 0.01$ &  $3.50 \pm 0.01$  & $21.68 \pm 0.02$   & $0.44 \pm 0.01$     &  $44.19 \pm 0.01$  &   $8.62 \pm 0.01$    &  $1$   &  $69.82 \pm 0.05$   & $0.45 \pm 0.01$      &  $43.42 \pm 0.02$    &   $9.37 \pm 0.02$    &  $1$  &   $30.55 \pm 0.02$  &  $0.22 \pm 0.01$    &  $44.83 \pm 0.02$   \\
Ks  &  $8.23 \pm 0.02$ &  $3.50 \pm 0.01$  & $21.68 \pm 0.02$   & $0.44 \pm 0.01$     &  $44.19 \pm 0.01$  &  $8.82 \pm 0.01$     &  $1$   &   $72.83 \pm 0.04$  & $0.45 \pm 0.01$      &  $43.42 \pm 0.02$    &   $9.93 \pm 0.03$    &  $1$  &  $33.12 \pm 0.02$   &  $0.22 \pm 0.01$    &  $44.83 \pm 0.02$   \\
3.4 & $8.33 \pm 0.02$  &  $3.50 \pm 0.01$  & $21.68 \pm 0.02$   & $0.44 \pm 0.01$     &  $44.19 \pm 0.01$  &    $8.73 \pm 0.01$   &  $1$   &  $79.93 \pm 0.04$   & $0.45 \pm 0.01$      &  $43.42 \pm 0.02$    &   $9.82 \pm 0.01$    &  $1$  &  $39.21 \pm 0.03$   &  $0.22 \pm 0.01$    &  $44.83 \pm 0.02$   \\
4.6 & $9.06 \pm 0.03$  &  $3.50 \pm 0.01$  & $21.68 \pm 0.02$   & $0.44 \pm 0.01$     &  $44.19 \pm 0.01$  &    $9.39 \pm 0.01$   &  $1$   &  $86.86 \pm 0.08$   & $0.45 \pm 0.01$      &  $43.42 \pm 0.02$    &  $10.22 \pm 0.01$     &  $1$  &  $45.16 \pm 0.04$   &  $0.22 \pm 0.01$    &  $44.83 \pm 0.02$    \\ \hline
\end{tabular}}
\label{table_results_model}
\end{table*}

\begin{table}
\caption{Magnitudes from best fit model of NGC 3115}
\centering
\begin{threeparttable}
\begin{tabular}{ccc}
\hline                                                  
Band & mag({\sc GALFITM}) & mag(NED) \\ \hline \hline
FUV* & $16.3 \pm 1.6 $ & $15.8 \pm 0.1$ \\
NUV & $14.3 \pm 1.4$ & $14.5 \pm 0.1$ \\
g & $10.0 \pm 1.0$& -- \\
r & $9.0 \pm 0.9$& $9.05 \pm 0.03$\\
i & $9.0 \pm 0.9$ & $8.72 \pm 0.03$\\
z & $8.7 \pm 0.9$ & -- \\
J & $6.7 \pm 0.7$ & $6.80 \pm 0.02$\\
H & $7.4 \pm 0.7$ & $7.9 \pm 0.02$\\
K & $7.6 \pm 0.8$ & $7.6 \pm 0.02$\\
3.4 & $7.6 \pm 0.8$ & -- \\
4.6 & $8.3 \pm 0.8$ & -- \\ \hline
\end{tabular}
\begin{tablenotes}
     \small
     \item *this magnitude was retrieved with the SS model.
    \end{tablenotes}
    \end{threeparttable}
\label{mag_results_model}
\end{table}

The shape and size of the components of a model can bring great insight to the actual morphology of a galaxy. For example, a classical bulge usually is modelled by a de Vancouleurs profile (i.e. S\'ersic indec (n) $\simeq$4), while a pseudo-bulge would have values of n $\simeq$ 2 \citep{Kormendy,Gadotti}). 
A thick disc would appear more oblate and extended than a thin disc, resulting in a higher value of b/a. Based on these considerations, we interpret our best-fit model of NGC 3115, composed of three components, as a representation of a galaxy containing a classical bulge, a thick disc and a thin disc.

Looking at the model images in Fig. \ref{model_BDD} and given the high value of the thick disc axis ratio (b/a=0.45) (see Table \ref{table_results_model}), this thick disc could be interpreted as an oblate spheroid as proposed by \cite{Arnold11} and \cite{Poci}. 
From Table \ref{table_results_model}, we see that we obtain a S\'ersic index of $n=3.5$ for the bulge, which is also in agreement with the black hole mass of NGC 3115, retrieved by \cite{Kormendy} ($2 \times 10^9$ M$_{\odot}$), according to \cite{Savorgnan} and \cite{Graham}.
Finally, in Table \ref{mag_results_model}, we compare the total magnitudes obtained in our {\sc GALFITM} model with available measurements in NED. We apply a conservative error of 10\% to our magnitudes, since this result is the combination of the magnitudes obtained for each component. We can see that, within errors, our {\sc GALFITM} model reproduces well the expected magnitudes.  The only exception is the total magnitude obtained for the FUV image using a BDD model, which is almost one magnitude brighter than the magnitudes available in the literature. The FUV image is too faint to fit three components and the model is overfitting the data, creating negative residuals, as possible to see in Fig. \ref{model_BDD}. In the rest of the work we will use the SS value for the FUV magnitude, when possible. In the other cases we will test the difference between including or not  the FUV value in the fit.

\begin{figure*}
\centering
    \includegraphics[width=\textwidth]{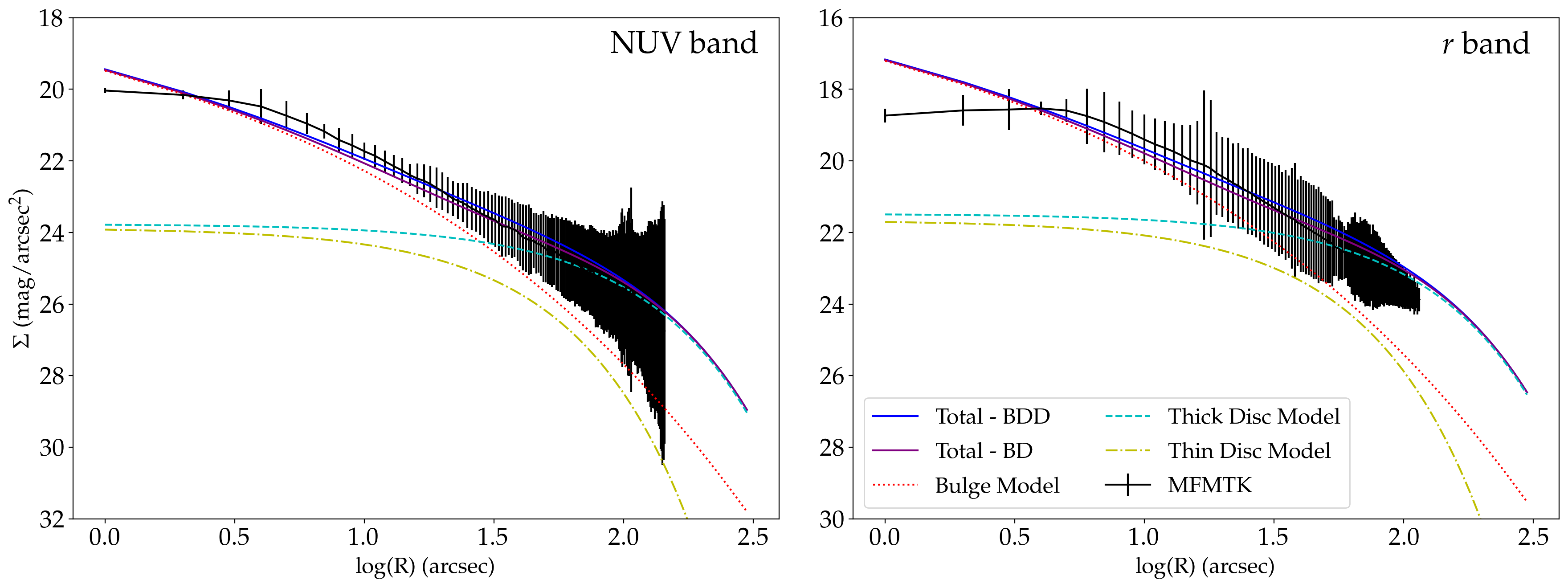}
    \caption{Surface density profiles of NGC 3115 and its components for two wavelength regimes: ultraviolet (NUV) and optical (r). Dotted red, cyan and yellow lines lines represent the extrapolation of the bulge (S\'ersic), thick disc (exponential disc), and thin disc (exponential disc) fit performed with {\sc GALFITM} to the stellar surface brightness data. The surface brightness profile as recovered with {\sc Morfometryka} (MFMTK) fitting (black dotted line) and with {\sc GALFITM} models with two (BD) and three (BDD) components (blue and magenta lines, respectively) overlap until the faint outskirts in NUV and from 60 arcsec in the optical.}
    \label{fig:profile}
\end{figure*}

In Fig. \ref{model_BDD}, we show the single S\'ersic, bulge-disc and bulge-disc-disc decomposition of NGC 3115. In it, we compare the accuracy of the fit when using one, two and three components. Firstly, we show in the right-hand side of the plot the median residuals over all bands, which consistently shows how the model with three components significantly improves the fit of the galaxy, achieving the smallest residual. We also show, for each band, the mean residual after the BDD decomposition, revealing that the bands that have the largest residuals are the ones in the optical, as it would be expected, since these are the deepest images, and as previously discussed, only in these bands we are able to identify and even fit extra components, such as a central bar (see Fig. \ref{residuals}) and spiral arms (see Fig. \ref{fig:ellipse_spiralarms}). The other bands are so faint that these extra components are not identified and so their residuals are smoother, since the main components of the galaxy are well fit. 

Fig. \ref{model_BDD} also shows each individual component fitted in the BDD model: the bulge, in the eighth row, the thick disc in the nineth row and the thin disc in the last row.
The values of best fit parameters for each band are shown in Table \ref{table_results_model}, as well as the order of the series of each parameter.

In Fig. \ref{fig:profile} we show the surface brightness profile of NGC 3115, highlighting where each component dominates the galaxy in two wavelength regimes: UV (NUV) and optical ($r$), as well as the profile recovered calculating the light encompassed in ellipses of fixed axial ratio obtained using {\sc Morfometryka} \citep{Ferrari}. 
One important comment on Fig. \ref{fig:profile} is that even if the modelled bulge is not dominant at large radii, it is a component that contributes from the innermost region of NGC 3115 until its outskirts, since by construction, the {\sc GALFITM} fit extends to infinite, becoming negligible where the galaxy light reaches the background value. As a consequence, the galaxy components also extend to infinite. The relative contribution of each component to the entire flux varies with radius and from galaxy to galaxy. \cite{Allen}, modelling the light distribution in 10095 galaxies from the Millennium Galaxy Catalogue (MGC), identified six types of surface brightness profiles. In the type 3 case, the S\'ersic profile accounts for the bulge light and the outer part light, being the disc embedded in the spheroidal component. The question whether the outer spheroid is simply an extension of the inner bulge or not, although interesting, is out of the scope of this paper. This method is different from other structural decomposition methods \citep[e.g.][]{Poci} that cut the domain of the bulge to a defined radius, and, therefore, are not directly comparable with this work.

The comparison with the light profile recovered with {\sc Morfometryka} shows that the total (bulge + thin disc + thick disc) {\sc GALFITM} model is recovering relatively well the galaxy light both in NUV and r bands, shown as an example. 
The displacement between {\sc GALFITM} and {\sc Morfometryka} at small radii can be explained by a S\'ersic core displacement \citep{Graham, Dullo}, this is only apparent with {\sc Morfometryka}, since this method fits isophotes to recover the galaxy light, while {\sc GALFITM} fits an extrapolation of the bulge S\'ersic profile.
However, an important difference between the two methods is that in the models recovered by {\sc GALFITM}, the axis ratio varies for different components, and the axial ratio of the total model is the sum of the contribution of the bulge, thin disc and thick disc, where the latter might vary with radius as well, depending on which component dominates the light profile. On the other hand, {\sc Morfometryka} calculates the light embedded in ellipses of fixed ellipticity. This difference does not seem to cause discrepancies, as the light profiles recovered with both methods are consistent within errors at all radii in the two wavelength regimes, probably as a consequence of the fact that the thick disc is the predominant component, in this radial range, and its axis ratio is the same obtained with {\sc Morfometryka}. With this test we show that the total parametric model of {\sc GALFITM} does not miss any light when compared with the flux enclosed in ellipses following the isophotal profile.

\subsection{The residuals}
\label{sec:equatorial}
The residuals of NGC 3115, i.e., the image obtained after subtracting the best fit model from the input images, are very peculiar and show the same patterns independently of the code used to fit the image (see \citealt{Guerou,Arnold11,Poci}), raising the question of whether there are other components in the galaxy. 

To clearly understand the residuals, we provide in Fig. \ref{evolution_residuals} the evolution of the residual images from the simplest {\sc GALFITM} model with one component to the one with three components for the $r$ band. By looking at Fig. \ref{evolution_residuals}, we begin to understand the necessity of a model with at least three components to describe NGC 3115. The left panels show the residual images resulting from fitting a single component: only the outer parts of the galaxy are fitted, leaving behind two semicircular symmetric regions in the major axis and a central equatorial component.  
In the two-components model (central panels of Fig. \ref{evolution_residuals}), the outer residuals around the major axis disappear, but the central equatorial component is still not fitted. Note that this equatorial component extends until nearly $1R_e$. Finally, adding a third component (right panels of Fig. \ref{evolution_residuals}), which we interpret as a thin disc, the model is capable of fitting a significant part of this equatorial component seen in both previous models. The fit of this third component reveals the existence of other two components on NGC 3115: two structures distributed along the major axis of the galaxy that can be either remnant of spiral arms or an edge-on ring; and a central hourglass shape, oriented along the minor axis of the galaxy, resembling an end-on bar. This model, moreover, presents the smallest median value of the residuals (when considering all the pixels in the residual image).
These findings are consistent with \cite{Capaccioli} and \cite{Guerou}, which proposed that the outskirts of the galaxy exhibit remnant spiral arms, which can be seen as very shallow symmetric structures in almost every residual image of the galaxy (see Fig. \ref{evolution_residuals}). \cite{Guerou} also proposes that the galaxy holds a bar and a nuclear ring in its centre.

As mentioned, analysing the residual image, we also observe a central component similar to a bar. Adding a fourth component to our {\sc GALFITM} fit (a S\'ersic model with S\'ersic index lower than 1, typical of a bar), we were able to fit such structure. However, we could only add this fourth component in the model of the Subaru images, because of the higher S/N (see Section \ref{sec:bar} for more details). 

It is hard to decouple the presence of the thin disc from a more general ``equatorial'' component, which would include not only the thin disc but also a nuclear ring, an inner bar, spiral-features, i.e. all the components lying along the central part of the major axis of the galaxy \citep{Arnold11,Guerou}. In an attempt to quantify the contribution of the equatorial component we obtained the flux of this inner region using the package {\sc Isophote} from the {\sc Photutils} \citep{photutils} library in Python, a tool to fit elliptical isophotes to a galaxy image, i.e. we fitted the total model - bulge - thick disc to quantify the flux of this inner region, and we compare it with the flux of the thin disc, as recovered by {\sc GALFITM}. The difference between the two is lower than 0.23, 0.16 and 0.27 magnitudes in the $g$, $r$ and $i$ bands, respectively (which is consistent with the errors in the {\sc GALFITM} model), so we conclude that the thin disc as extracted with {\sc GALFITM} is a fair representation of the central region of the galaxy. In Section \textsection \ref{sec:discussion} we discuss in detail the result of the ellipse fitting for the different sub-components.

\section{Analysis}
\label{sec:analysis}

\subsection{Colours}
\label{sec:colors}
\begin{figure}
    \includegraphics[width=\columnwidth]{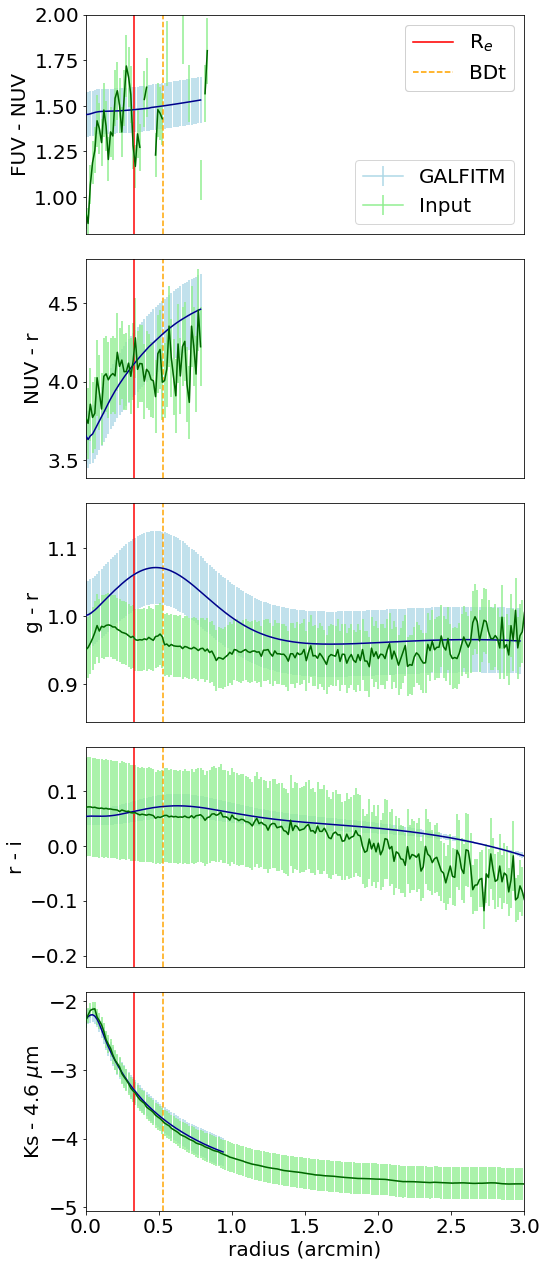}
    \caption{Colour across semi-major axis for NGC 3115, using colours in different wavelength ranges. UV: FUV-NUV and NUV-r, optical: g-r and r-i, near-IR: Ks-4.6 $\mu$m. The dark blue line reflects the {\sc GALFITM} model colour measurements in each pixel from the centre to the outermost region of the galaxy, with shaded errors corresponding to 3$\sigma$, while the dark green line shows the colour gradient retrieved using the input images. The red vertical line highlights the place where the bulge ends according to our {\sc GALFITM} model (Re = 21.7 arcsec), the yellow vertical line shows the bulge-to-disc transition recovered from Fig. \ref{fig:profile}.}
    \label{fig:1Dcolor_gradient}
\end{figure}

Our first step to understanding the stellar populations present across the galaxy is deriving 1D colour gradients of our models.
In Fig. \ref{fig:1Dcolor_gradient} we show different colours of the galaxy retrieved from the {\sc GALFITM} models and input images, i.e. FUV-NUV, NUV-$r$, $g$-$r$, $r$-$i$ and Ks-4.6.

To create this figure we draw a rectangle of 50 arcsecs height along the galaxy's major axis, from the innermost pixel of NGC 3115 until its outskirts, covering out to 3 arcmin, and we obtain the median value of the colour at each radius (in bins of 1 arcsec). In this figure, the red vertical line represents the end of the bulge domain in the r band (21.69 arcsec) as derived from the bulge model of {\sc GALFITM} (see Table \ref{table_results_model}), and the yellow vertical line shows the bulge-to-disc transition (i.e. where the bulge light becomes subdominant in the fit) derived from Fig. \ref{fig:profile}.

To create this figure, we use both the {\sc GALFITM} models (blue line)  and the input images (green line), for comparison.

In Fig. \ref{fig:1Dcolor_gradient} we show the colour gradients obtained for the two cases and we see that we recover, within errors, the same trends using the input image and modelled image, indicating that we are retrieving reliable colour gradients present in the galaxy. From Fig. \ref{fig:1Dcolor_gradient}, we can observe, moreover, the power of {\sc GALFITM} in delivering a  smoother colour profile, and enhancing regions of transition and inversion of the colour gradient.

It is worth noticing in Fig. \ref{fig:1Dcolor_gradient} that the region that demarcates the transition from the bulge to the disc domain (extracted from Fig. \ref{fig:profile}) coincides with the inflexion in the colour gradient in the optical bands, suggesting the presence of different stellar populations between the components. Moreover, it is interesting to see that the galaxy becomes bluer with radius in the optical, indicating that we may be observing the effect of several low-mass accretions in NGC 3115 that would account for the bluer colours in the outskirts of the galaxy \citep{Mapelli}.

Moreover, from the {\sc GALFITM} model gradients in Fig. \ref{fig:1Dcolor_gradient}, it is clear that in the ultraviolet range the inner region is bluer than the outer one. Similar behaviour is present in the optical colour, but it is inverted at large radii, being the transition point the radius where the bulge light gets subdominant. The near-IR colour gradient is opposite to the FUV-NUV and NUV-r ones, being the inner region the reddest.

According to \cite{Kaviraj}, a recovered NUV - r $\simeq 3.3$ colour would imply that the bulge of this galaxy hosts a young stellar population, as discussed in details in Section \ref{sec:formation}.
Finally, the optical colours $g-r$ and $r-i$ are consistent with the results obtained by \cite{Guerou, Poci, Dolfi}:  from the outskirts inwards, we detect a bluer outer envelope, followed by a redder disc-like region and by a mid-colour inner region.

The bluer colour of the bulge in the ultraviolet regime could be attributed to a young stellar population, the presence of an AGN \citep{Wong} or the emission of white dwarf stars \citep{Lisker}.
However, the latter is less likely, because white dwarfs would yield bluer FUV-NUV, instead of NUV-r, and the contribution of white-dwarfs emission would not be enough to explain alone the retrieved colour gradients in the FUV-NUV and NUV-r images. 
Moreover, this galaxy presents extended strong emission in the Spitzer $8$ and $24$ micron bands, covering especially the central part of the galaxy.
The presence of the mid-infrared emission in passive early-type galaxies (ETGs) might be of stellar origin, coming from the dusty envelopes around the asymptotic giant branch (AGB) stars \citep{Clemens} or could, again, indicate the presence of an AGN.
In fact, there is evidence for the presence of a billion mass supermassive black hole ($2 \times 10^9$ M$_{\odot}$) in this galaxy and hot flows in the centre, studied in detail by \cite{Jones,Almeida,Kormendy96,Lauer,Emsellem}.

By looking again at Fig. \ref{fig:1Dcolor_gradient}, we see that using optical and near-IR colours, we retrieve the same gradient as the one found by \cite{Guerou}, i.e. a bulge with an older/more metallic stellar population than the disc. However, when we include ultraviolet colours, we can identify these blue colours in the central region of the galaxy. Several pieces of evidence strengthen this hypothesis:
(1) The very central region of the galaxy (identified as blue in the $NUV-r$ panel) is the only one that shows X-ray emission. (2) This same region shows high dust temperature when observed with Spitzer \citep{Amblard}, indicating that the UV light was likely absorbed and re-emitted by the dust, confirming the presence of strong UV light in this region (strong evidence of young populations), and (3) when focusing only on optical and near-IR colours, our results agree well with previous works. 

To understand the origin of the blue colour in the central region of the galaxy, i.e. if it is due to the presence of an AGN or recent star formation activity, we perform a fit of the SED of the galaxy and its components, as described in the next section.

\subsection{SED fitting}
\label{sec:sed}
In this section, we describe the SED fitting process implied to derive the physical properties of the galaxy and its components.
For such task, we used the code {\sc CIGALE} \citep{Burgarella,Noll}.

\subsubsection{\sc{CIGALE}}

{\sc CIGALE} is a robust SED fitting code, which includes the contribution of dust emission and attenuation models, AGN models, as well as the simple stellar population models from \cite{BC03}, providing accurate estimates of the galaxy properties.

Since we aim to understand if the blue colour observed in the bulge is the result of recent star formation events in this region or emission of the AGN (or a composite of the two), we use {\sc CIGALE} to perform the fitting accounting for models that include both the contribution of stellar and AGN emission. 

We fit a total of 11 free parameters to the SED of the galaxy and its components: 4 parameters for the star formation history (SFH), 1 parameter for the dust attenuation, 1 parameter for the simple stellar population and 5 parameters for the AGN emission templates, described in Table \ref{table:cigale_params}. 

In the SED fitting process, we include the AGN emission models of \cite{Fritz} only in the bulge component and we use the stellar population models of \cite{BC03}. We assume the dust attenuation model described by \cite{Calzetti}, a Chabrier initial mass function \citep{Chabrier}, and a double exponential SFH, which, according to \cite{Burgarella}, provides a good estimate of the SFR of early-type galaxies and galaxies that had a recent episode of star-formation, but is not good for recently quenched or starburst galaxies. The concept behind this star formation history is to allow for two decaying exponentials to fit the star formation rate of the system, one for the long term SFR and one for the more recent events. We point out, however, that it is not because we are using a double exponential SFH that both peaks will have significant participation in the fit. In fact, if the data requires, one of the peaks of the SFH has the freedom to be zero.

It is especially important to include AGN emission in the SED fitting of the bulge component, as it can have a large impact on the derived properties of the component and bring light to the role of the AGN in the evolution of lenticular galaxies. For example, in the UV and optical wavelength range, a large portion of the emitted light of a galaxy hosting an AGN can be attributed to the accretion disc, and could thus explain the blue colours that are observed in the bulge of NGC 3115.

\begin{table}
\caption{Parameters used in the SED fitting procedure with {\sc CIGALE}}
    \scalebox{0.9}{%
\begin{tabular}{@{}lc@{}} 
\toprule
Parameter & \textbf{Value} \\ 
\midrule \midrule
 & \textbf{Double exp. decreasing SFH} \\ 
\midrule
$\tau_{main}$ (Gyr) & 9,10,11,12,13 \\
$f_{burst}$ & 0.20\\
$age_{burst}$ (Gyr) & 0.1,1,2\\
$\tau_{burst}$ (Gyr) & 1,5\\ \hline
 & \textbf{Simple Stellar Population - BC03} \\ \hline
IMF & Chabrier \citep{Chabrier}\\
Metallicity [Z/H] & 0.008, 0.02,0.05 \\ \hline
 & \textbf{Dust attenuation} \\ \hline
$A_{V}$ (mag) & 0.50, 1.0, 2.0, 3.0, 5.0\\ \hline
 & \textbf{AGN \citep{Fritz}} \\ \hline
$\tau$ & 0.1,0.3 \\
$\beta$ & -0.75, -0.50, -0.25, 0.00\\
$\gamma$ & 0.0, 2.0, 4.0, 6.0\\
Opening angle (deg) & 60., 100., 140\\
AGN fraction & 0.001,0.1 \\ \hline
\# of models - without AGN & 2160\\ 
\# of models - with AGN & 829440 \\ 
\bottomrule
\end{tabular}}
\label{table:cigale_params}
\end{table}

In Table \ref{table:cigale_params} the fitted parameters and the values each parameter can assume in a {\sc CIGALE} SED modelling are shown. The total number of models, shown in the bottom line of Table \ref{table:cigale_params}, states the total number of SED models that are created by varying and combining these parameters. The parameters described in Table \ref{table:cigale_params} are: $\tau_{\rm main}$, the e-folding time of the main stellar population model, $f_{\rm burst}$ is the mass fraction of the late burst population, ${\rm age}_{\rm burst}$ is the age of the late burst and $\tau_{\rm burst}$ is the e-folding time of the late starburst population model. While for the dust attenuation the only fitted parameter is $A_v$, the V-band attenuation of the young population. For the AGN, $\tau$ stand for the optical depth at 9.7 microns, while $\beta$ is linked to the radial dust distribution in the torus and $\gamma$ is linked to the angular dust distribution in the torus. The opening angle regards the full angle of the dust torus (see Fig. 1 of \citet{Fritz}), and the AGN fraction relates to the percentage of the AGN mass concerning the total mass of the galaxy.

To find the best fitting SED, we performed millions of tests using {\sc CIGALE}, using both the delayed SFH and double exponential SFH. What we see is that both models fit well the data, however, the delayed SFH fails to reproduce simultaneously the UV (young stellar population) with the NIR and IR (old stellar populations), delivering sometimes too high SFR for very old stellar populations. As discussed in \cite{Ciesla16,Ciesla17}, it is very hard to recover both age and SFR reliably using SED fitting techniques. Moreover, there is a general agreement on the difficulty of constraining the age of galaxies from broad-band SED fitting \citep[e.g.][]{Ciesla15, Pforr12, Maraston10, Buat14}. What we find, however, is that using a double exponential SFH, the delivered ages are closer to the ones expected from other works while maintaining the physical reliability of the SFR.

Therefore, we decided to use the double exponential SFH for all SEDs shown in this work.

\begin{figure*}
\centering
\includegraphics[width=\textwidth]{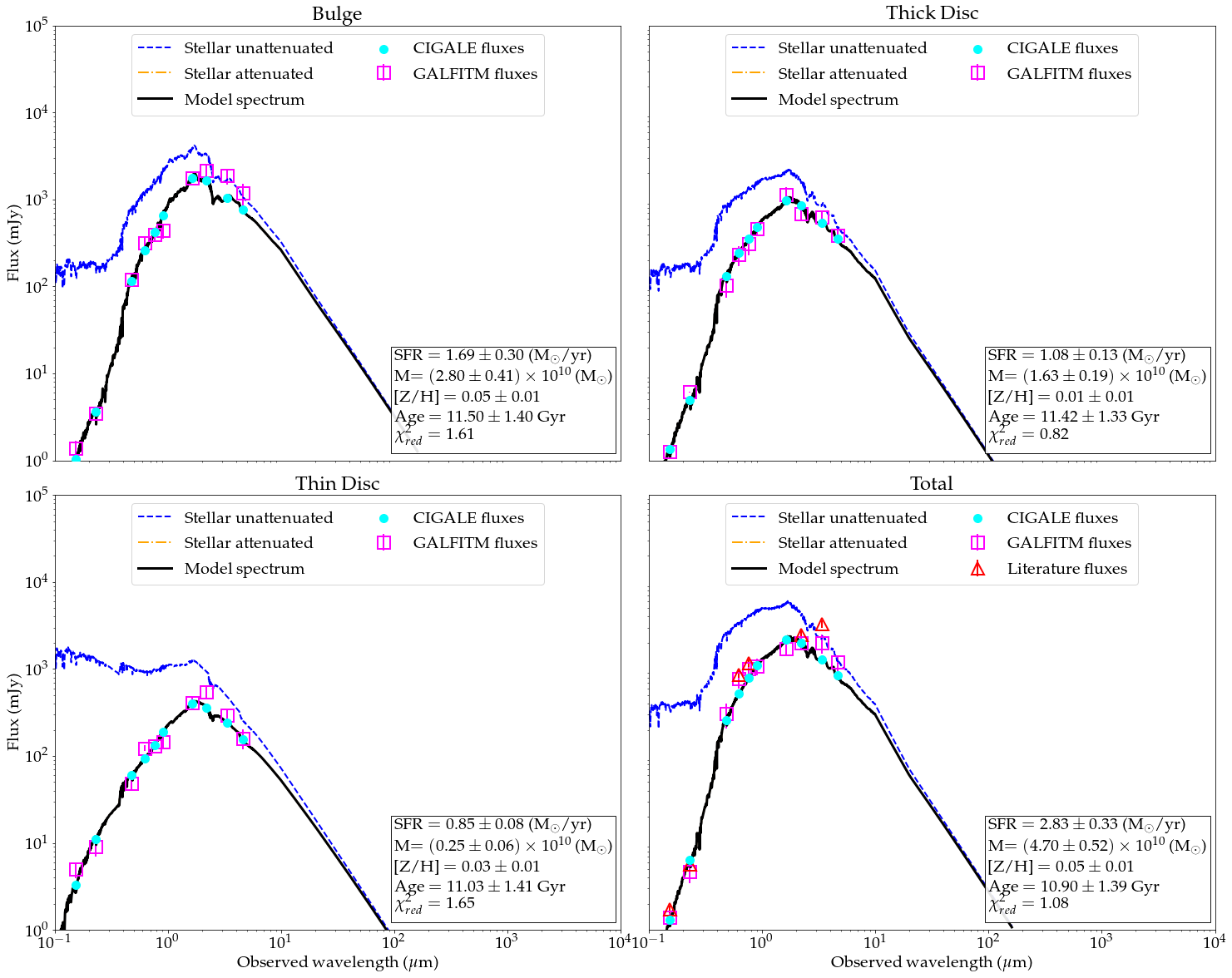}
\caption{Spectral energy distribution of each subcomponent of NGC 3115: bulge, thick disc and thin disc, and the total model of the galaxy, respectively. In each panel, we show the retrieved physical properties. The blue squares stand for the input fluxes, the red dots are the model fluxes, the black line is the modelled spectrum, the orange line is the stellar attenuation, the red line is the dust emission and the blue dashed line is the stellar emission unattenuated.}
\label{fig:SED_withoutAGN}
\end{figure*}

\begin{figure}
\centering
\includegraphics[width=\columnwidth]{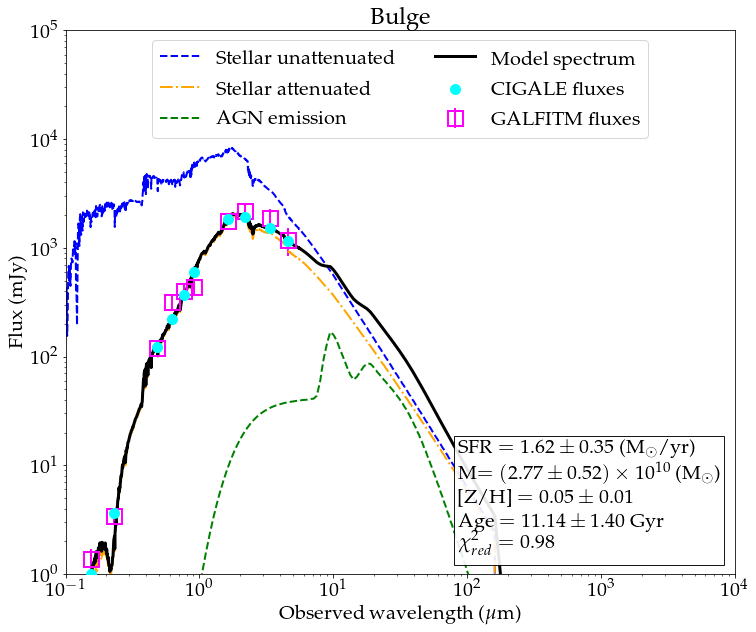}
\caption{Spectral energy distribution of the bulge of NGC 3115 including AGN models. The blue squares stand for the input fluxes, the red dots are the model fluxes, the black line is the modelled spectrum, the orange line is the stellar attenuation, the red line is the dust emission, blue dashed line is the stellar emission unattenuated and green line is the AGN emission.}
\label{fig:SED_withAGN}
\end{figure}

The resulting SEDs of the galaxy and its components are shown in Fig. \ref{fig:SED_withoutAGN}, while the SED of the bulge including AGN models is shown in Fig. \ref{fig:SED_withAGN}. In these figures, we also indicate the contribution of dust attenuation and stellar emission. 

\begin{figure*}
\centering
\includegraphics[width=\textwidth]{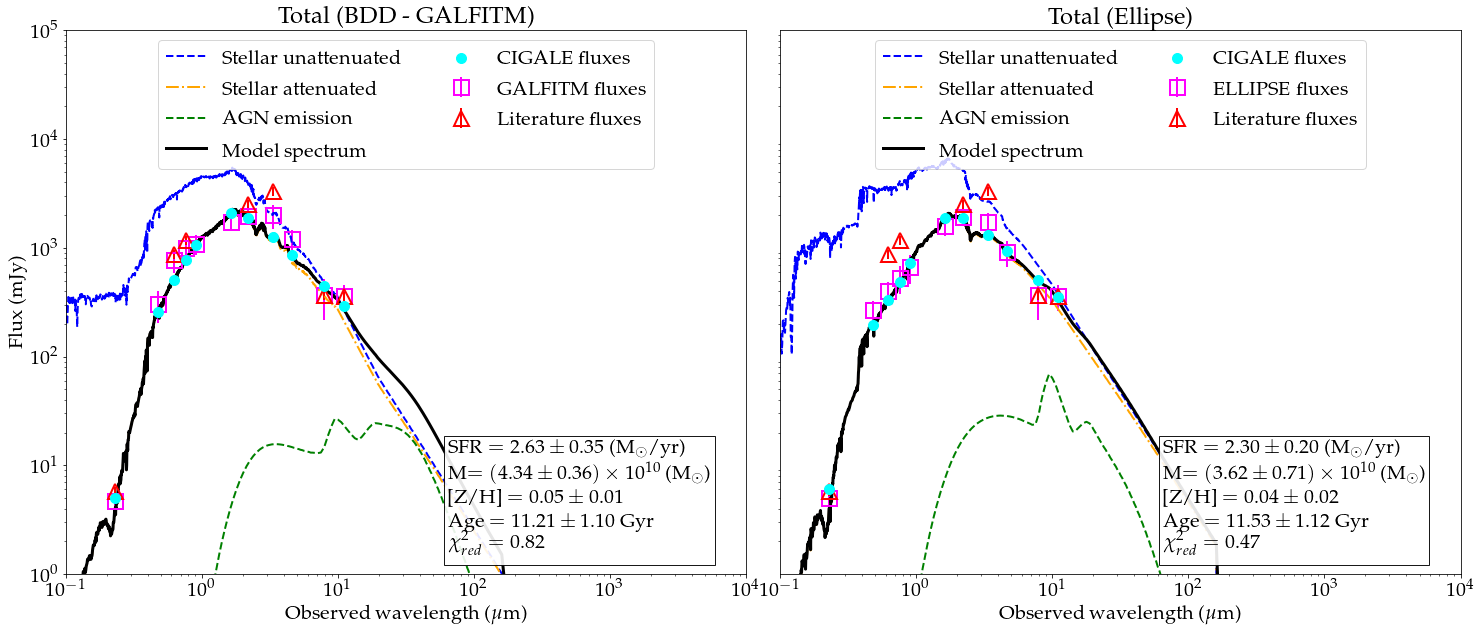}
\caption{Spectral energy distribution of the NGC 3115 photometry obtained using {\sc GALFITM} and {\sc ELLIPSE}, respectively. The first eleven points in our fit represent the data used throughout this work, while the final two points are literature data from Spitzer 8$\mu$ and IRAS 12$\mu$. In each panel, we show the retrieved physical properties. The blue squares stand for the input fluxes, the red dots are the model fluxes, the black line is the modelled spectrum, the orange line is the stellar attenuation, the red line is the dust emission, blue dashed line is the stellar emission unattenuated and green line is the AGN emission.}
\label{fig:SED_bddellipse}
\end{figure*}

Furthermore, to better constrain the SED results and be certain we are obtaining fiducial results, we took two complementary roads: (1) we added to our fitting procedure two literature data in the infrared regime \citep[Spitzer 8$\mu$, IRAS 12$\mu$][]{Amblard,Knapp}. (2) we compare the results obtained with {\sc GALFITM} with those obtained with {\sc ELLIPSE} in {\sc IRAF}. The results of this comparison, with the infrared complementary data included in the fitting, are shown in Fig. \ref{fig:SED_bddellipse}.

The results presented in Figs. \ref{fig:SED_withoutAGN}, \ref{fig:SED_withAGN} and \ref{fig:SED_bddellipse} are summarised in Table \ref{table:sed_cigale}, showing the main physical properties obtained for the galaxy and its components, including and not including AGN models in the fitting process. In the two bottom rows, we show the results when including literature data to constrain the SED fitting and compare it with the fit obtained with {\sc ELLIPSE} in {\sc IRAF}.

\begin{table*}
\caption{SED Fitting results obtained from {\sc CIGALE}. The properties obtained with {\sc CIGALE} are divided in three: containing AGN models in the fitting process, without AGN models, and including literature infrared data.}
    \hspace*{-0.5cm}
    \ra{1.2}
    \scalebox{1.0}{%
\begin{tabular}{@{}lcccccc@{}} 
\toprule
 & \textbf{Component} &  \textbf{SFR (M$_{\odot}\, $yr$^{-1}$)} & \textbf{Z} & \textbf{M$_{\star} \, ($M$_{\odot}$)} & \textbf{sSFR (yr$^{-1}$)} & \textbf{Age (Gyr)}\\ 
\midrule \midrule
\textbf{With AGN} & Bulge 		&	$1.6 \pm 0.4$                  &  $0.05 \pm 0.01$        & $(2.8 \pm 0.5) \times 10^{10}$  & $(7.1 \pm 2.2) \times 10^{-11}$ & $11.2 \pm 1.4$   \\ \hline
 \multirow{4}{*}{\textbf{Without AGN}} & Bulge 		&	  $1.7 \pm 0.3$                    & $ 0.05 \pm 0.01 $       & $(2.8 \pm 0.4) \times 10^{10} $ &  $(6.0 \pm 1.4) \times 10^{-11}$   & $11.5 \pm 1.4$   \\ 
 & Thin disc 	&	 $0.8 \pm 0.1$                 & $0.03 \pm 0.01 $       & $(0.3 \pm 0.1) \times 10^{10} $ &   $(2.7 \pm 1.1) \times 10^{-10}$  & $11.0 \pm 1.4$    \\ 
 & Thick disc 	&	  $1.1 \pm 0.1$                    & $ 0.01 \pm 0.01$        & $(1.6 \pm 0.2) \times 10^{10}$  &  $(9.2 \pm 2.4) \times 10^{-11}$  & $11.4 \pm 1.3$  \\ 
 & Total              & $2.8 \pm 0.3$                   & $0.05 \pm 0.01$       & $(4.7 \pm 0.5) \times 10^{10}$  & $(6.0 \pm 0.9) \times 10^{-11}$  & $10.9 \pm 1.4$   \\ \hline
  \multirow{2}{*}{\textbf{With IR}} & Total 		&	$2.6 \pm 0.3$                  &  $0.05 \pm 0.01$        & $(4.3 \pm 0.4) \times 10^{10}$  &  $(6.0 \pm 0.9) \times 10^{-11}$  & $11.2 \pm 1.1$   \\
  & {\sc ELLIPSE} 	&	       $2.3 \pm 0.2$               & $0.04 \pm 0.02$        & $(3.6 \pm 0.7) \times 10^{10}$     &  $(6.4 \pm 1.3) \times 10^{-11}$  & $11.5 \pm 1.1$               \\ 
\bottomrule
\end{tabular}}
\label{table:sed_cigale}
\end{table*}

From Figs. \ref{fig:SED_withoutAGN} and \ref{fig:SED_withAGN} and Table \ref{table:sed_cigale}, we begin to understand the influence of the AGN in the emission of the galaxy bulge at different wavelength ranges. As other works have shown before \citep{Dolfi, Guerou}, the galaxy is overall old and with low amounts of gas in all components. However, using photometric data in the UV, we detect blue colours in the bulge of the galaxy (\ref{sec:colors}), maybe revealing the presence of younger/less metallic populations in its central region. The emergence of this blue colour could be explained by a recent star formation event in the centre of the galaxy, or it could be due to AGN emission. What we see, however, in Fig. \ref{fig:SED_withAGN}, is that the AGN emission has very low luminosity \citep{Jones, Almeida} and it is always subdominant in the galaxy, especially in the UV regime, where this emission seems not to play any role. This indicates that the bluer colours observed in the bulge can not be due to the presence of the AGN. Moreover, looking at Table \ref{table:sed_cigale} we see that the derived properties obtained for the bulge including and not including AGN models are similar, within errors, reinforcing that the presence of this AGN is not interfering in the physical properties of the bulge. Note that the reduced $\chi^2$ is smaller in the SED fitting including the AGN model.

From Table \ref{table:sed_cigale}, we see that the mass of the galaxy is mainly in the bulge, where the supermassive black hole lies, in contrast with what was found by \citet{Poci} that the majority of the galaxy's mass is in the spheroidal/thick disc component. This discrepancy might be related to the different methods used to recover the galaxy's components. In fact, in \citet{Poci}, an orbit-superposition dynamical modelling is used, while {\sc GALFITM} fits the galaxy light distribution. This implies that the S\'ersic model used to recover the bulge component would be dominant in the innermost region, but would keep a contribution until very large radii, resulting in the high mass recovered for the bulge component. On the other hand, \citet{Poci} associate all the outer part of the galaxy to the spheroidal/thick disc component. 

To investigate the validity of the decomposition carried out in this work, we calculate the mass-to-light (M/L) ratio for the bulge and thick disc components using the $3.4\mu$m band. The results obtained are comparable to what found by the DiskMass Survey \citep[DMS;][]{Bershady,Lelli}, being  ${\rm M/L}_{\rm bulge} = 0.26$ for the bulge and  ${\rm M/L}_{\rm disk} = 0.16$ for the disk.
Moreover, we find that ${\rm M/L}_{\rm bulge} = 1.6\,\, {\rm M/L}_{\rm disk}$, nearly as suggested by SPS models \citep{Schombert}, granting the feasibility of the recovered parameters. Thus, we conclude that our decomposition is valid. Interestingly, \cite{Lelli}, for Spitzer data at $3.6\mu$m, find that ${\rm M/L}_{\rm bulge} = 0.7$, ${\rm M/L}_{\rm disc}= 0.5$. Keeping the light fixed, these values of M/L would imply almost double the mass recovered with SED fitting in this work, so nearly the same mass obtained by \citet{Guerou,Poci}.

In fact, when comparing the total stellar mass of the galaxy with the results obtained using MUSE data \citep{Guerou,Poci}, we can see that our measurement is lower, even using a bigger field-of-view. In this respect, it has been shown in several works (e.g. \cite{Ciesla15} and references therein) that SED fitting techniques deliver stellar masses systematically lower than their spectroscopic counterparts. Particularly, using {\sc CIGALE},  \cite{Buat14} and \cite{Ciesla15} found that the output stellar mass, independently of the SFH chosen, is approximately 7\% lower than expected.

When comparing the results obtained with {\sc GALFITM} for the total model of NGC 3115 (bulge+thick+thin disc) with results obtained with {\sc ELLIPSE} in {\sc IRAF}, we see that the results we are obtaining with {\sc GALFITM} do reflect realistic features of the galaxy. One important reinforcement to make in this comparison is that {\sc ELLIPSE} integrates the light of the galaxy until the outermost detectable (with respect to the background light) isophote, whilst {\sc GALFITM} integrates until infinite, so the magnitudes obtained are expected to be slightly fainter for {\sc ELLIPSE}, and the SED results might reflect these differences. Moreover, while the fitting with {\sc ELLIPSE} integrates the flux inside several ellipses, {\sc GALFITM} recovers the flux of the parametric models that better recover, when combining all the components, the light distribution of the galaxy. Thus, this comparison is only valid in order of magnitude, as a consistency check. Looking at Table \ref{table:sed_cigale}, it is possible to see that the total physical parameters obtained with the two models are in agreement, inside error bars.   
The results of stellar mass obtained using both {\sc GALFITM} and {\sc ELLIPSE} are consistent with the ones obtained by \cite{Forbed17b} using Spitzer imaging.

Therefore, excluding the mass, the stellar population properties recovered in this work are in agreement with what found by \citet{Guerou} using MUSE data.
One should keep in mind, nonetheless, that the two methods are not directly comparable\footnote{IFU data allow to study pixel-by-pixel the variation of the parameters for the whole galaxy, while GALFITM recovers averaged properties for each component}; for instance, it has been shown that SED fitting techniques tend to underestimate the recovered ages \citep{Ciesla}. Nevertheless, both works find that the stellar population of the galaxy is on average around $\simeq 11$ Gyr, with slight variation along the major axis, being the innermost part the oldest and more metallic, followed by a region with a mixed population (young and star-forming/old and quenched) and a middle-aged, low metallicity outer region. 

Finally, using the recovered stellar mass and SFR, we obtained the specific star formation rate (sSFR) for each component. We find that the thin disc has sSFR of around $2.7 \times 10^{-10}$ $yr^{-1}$ and it is higher than the sSFR of the thick disc and the bulge, see Table \ref{table:sed_cigale}. Following the definition of \cite{Bruce12} (sSFR $<$ $10^{-10}$ yr$^{-1}$ is quiescent), the thick disc and the bulge are quenched, while the thin disc is moderately active, again in agreement with what found by \citet{Guerou}. 
Interestingly, even if quenched, the bulge presents a recent event of star formation, as we shall discuss in detail in the next Section.

\section{Discussion}
\label{sec:discussion}

\subsection{The inner bar}
\label{sec:bar}
As mentioned in \textsection \ref{sec:equatorial}, the model in the optical bands is significantly improved with the addition of a bar, however, the inclusion of this bar to other bands with lower S/N caused a strong over subtraction and so we decided not to include this component in the final model. Looking for alternative ways to prove the hypothesis of the presence of this bar, we compared NGC 3115 with two simulated galaxies: both with the same fitted components as NGC 3115 (bulge, thick and thin disc), but one with a bar and one without a bar to see which had more similarities (and showed similar residuals to those) of the model.

The unbarred simulation uses the star-forming simulations described in \cite{Clarke} and \cite{ BeraldoSilva} (who refer to it as FB10), which forms clumps at early times producing a thick disc in the system. 
The barred galaxy, on the other hand, is a pure N-body (no gas or star formation) simulation initially comprised of a thin disc$+$thick disc. This model is similar to the thin disc$+$thick disc simulations described in \cite{debattista}, but with the initial thick disc having a scale-height of $900 pc$. After forming a bar, we view the system edge-on with the bar end-on.

For the comparison, we have created unsharp masks for the $r$ band of NGC 3115, the unbarred galaxy and the barred galaxy. The unsharp mask is an image sharpening technique, which consists of digitally blurring or ``smoothing'' the original image. This operation suppresses smooth features (i.e., which have a structure on large scales) in favour of sharp features (those with a structure on small scale), resulting in a net enhancement of the contrast of fine structure in the image. The unsharp masks used in this work were created using a 25-by-25 pixels median box to create the median subtracted image and a circular Gaussian smoothing with $\sigma = 5$.

\begin{figure}
\centering
\includegraphics[width=\columnwidth]{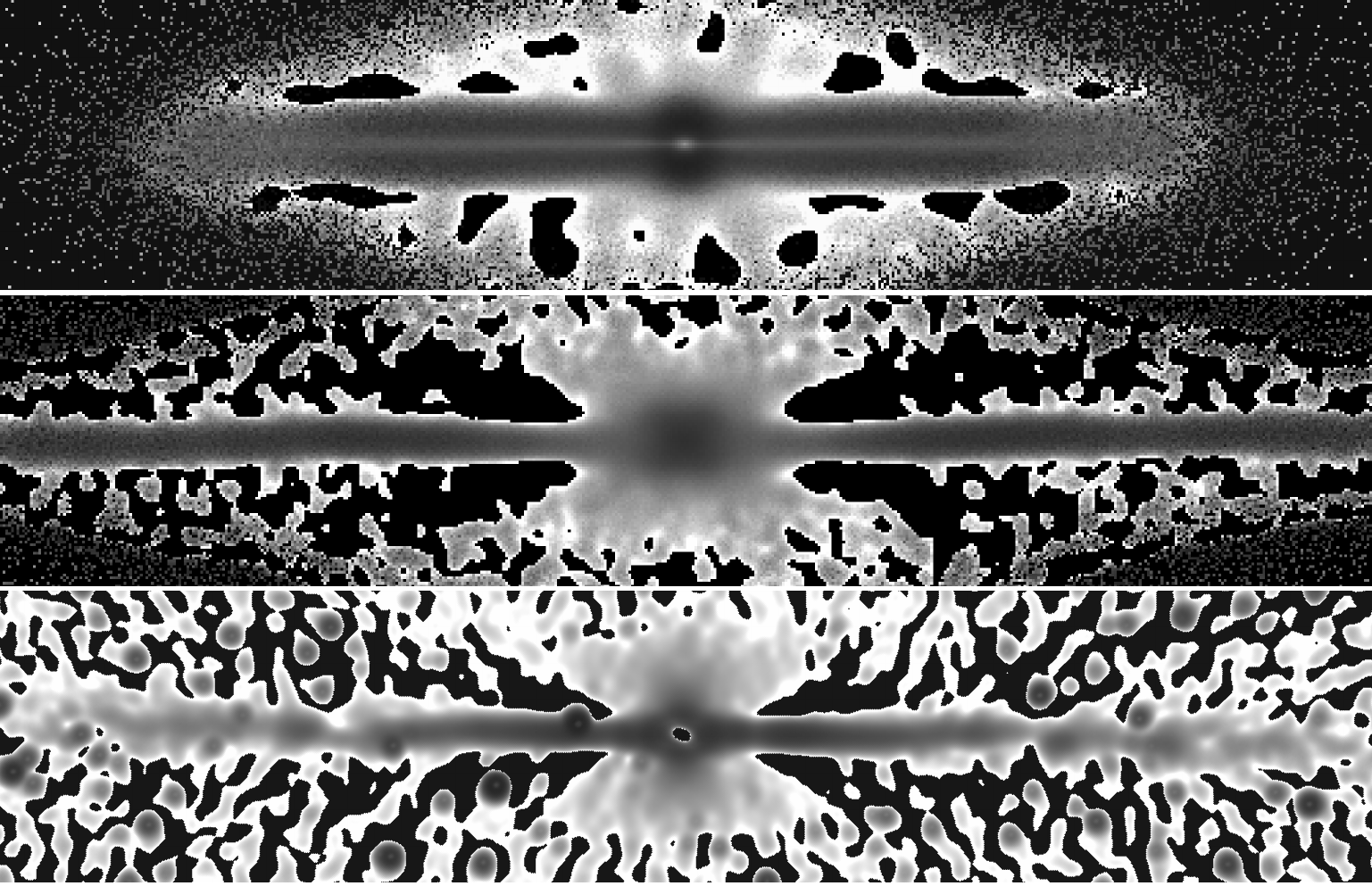}
\caption{\textbf{Top}, unsharp mask of a simulated unbarred galaxy with a bulge, thick and thin discs. \textbf{Centre}, unsharp mask of a simulated barred galaxy, also with a bulge, thick and thin discs. \textbf{Bottom}, unsharp mask of the r band of NGC 3115.}
\label{residuals}
\end{figure}

The top panel of Fig. \ref{residuals} shows the unsharp mask of the simulated unbarred galaxy, where it is possible to see a stripe of stars in the centre of the galaxy. The simulated barred galaxy is shown in the central panel of Fig. \ref{residuals} and the unsharp mask of the $r$ band of NGC 3115 in the bottom panel of Fig. \ref{residuals}.
It is clear that the hourglass shape present in the unsharp-mask of the $r$-band image is also present in the simulation of the barred galaxy, suggesting that this feature in NGC 3115 is probably caused by the presence of a bar. \cite{Guerou} also reported the possible existence of an end-on bar, based on the correlation between the mean and skewness of the measured line-of-sight velocity distribution (V and h3), usual of barred-galaxies.

\subsection{Recent dynamical history: encounter with a dwarf companion}
\label{sec:dwarf}
\begin{figure*}
    \centering
    \includegraphics[width=\textwidth]{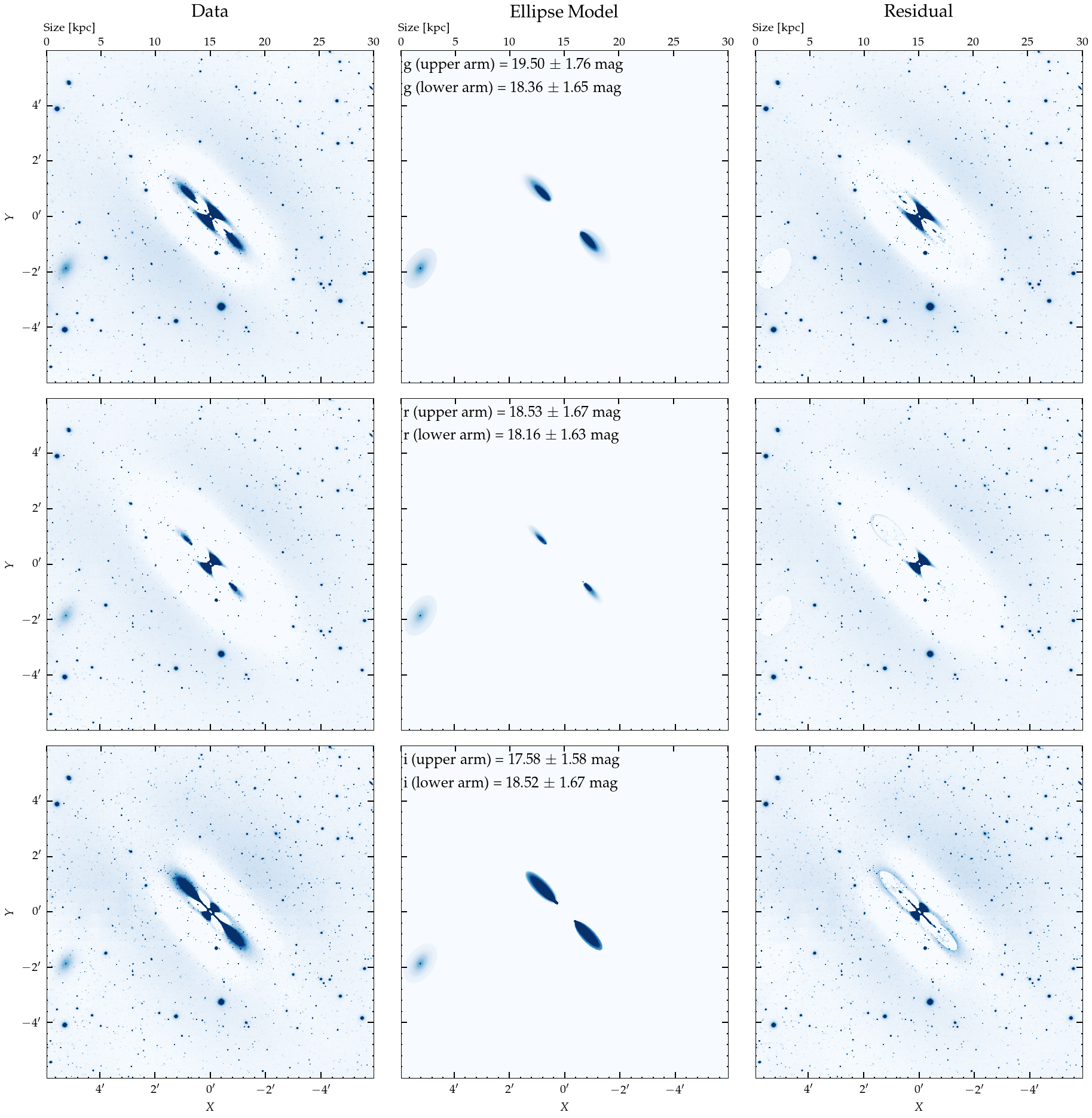}
    \caption{Ellipse fitting of the spiral-like features observed in NGC 3115 and its dwarf companion KK084, located approximately at (5',-2'). The rows represent the $g$, $r$ and $i$ bands, respectively. The columns show the input image (residual from the fit with {\sc GALFITM}), the model of the spiral features and companion KK084 and residuals of the fit, respectively.}
    \label{fig:ellipse_spiralarms}
\end{figure*}

NGC 3115 is known to have a very rich globular cluster system \citep{Cantiello} and several companion galaxies, which are mostly ultra-compact dwarfs \citep{Dolfi}. 
KK 084 is one of the dwarf spheroidal companions of NGC 3115, located at about 5.5 arcmin to the SE, \citep{Harris, Puzia}. KK084 is very faint and, from all datasets available to us, it could only be detected in the deep Subaru images.
Using these images, we fitted KK 084 with {\sc ELLIPSE} in the {\sc Photutils} library in Python, to recover the main physical properties of this dwarf galaxy. 
Also with {\sc ELLIPSE}, we fitted the observed spiral-like features in the residuals of the models obtained with {\sc GALFITM}. In total, we fitted three components with {\sc ELLIPSE}, the companion and the NE and SW spiral arms. The results of this fit are shown in Fig. \ref{fig:ellipse_spiralarms} and in Table \ref{table:companion}.

Analysing Fig. \ref{fig:ellipse_spiralarms}, it is intriguing how the colours of the two hypothetical remnants (or reborn) spiral arms differ from each other. This difference may be due to the very low S/N of the images of both structures, or, given that one arm is much redder than the other, it could be due to the presence of dust. However, only new deep multi-colour imaging ($\mu_r \simeq 28$) could provide a robust understanding of the stellar populations present in these sub-components.

\begin{table}
\caption[colours of NGC 3115 and KK084]{colour of the components of NGC 3115 and the dwarf companion KK084 recovered using {\sc GALFITM} and {\sc ELLIPSE}.}
    \hspace*{-0.5cm}
    \ra{1.2}
    \scalebox{0.9}{%
\begin{tabular}{@{}lccc@{}} 
\toprule
\textbf{Component} &  \textbf{g - r} & \textbf{g - i} & \textbf{r - i} \\ 
\midrule \midrule
 Total Model ({\sc GALFITM}) & $0.99 \pm 0.12$ & $1.05 \pm 0.14$ & $0.05 \pm0.01 $ \\
 Bulge ({\sc GALFITM}) & $0.95 \pm 0.13$ & $1.03 \pm 0.13$ & $0.08 \pm 0.02$\\
 Thick disc ({\sc GALFITM}) & $0.95 \pm 0.16$ & $0.99 \pm 0.14$ & $0.05 \pm 0.01$\\
 Thin disc ({\sc GALFITM}) & $1.11 \pm 0.14$ & $1.17 \pm 0.19$ & $0.06 \pm 0.02$\\
 KK084 ({\sc Ellipse}) & $0.63 \pm 0.07$ & $0.68 \pm 0.09$ & $0.05 \pm 0.01$\\
NE spiral feature ({\sc Ellipse}) & $0.93 \pm 0.14$ & $1.92 \pm 0.34$ & $0.77 \pm 0.11$\\
SW spiral feature ({\sc Ellipse}) & $0.2 \pm 0.01$ & $-0.16 \pm 0.03$ & $-0.36 \pm 0.07$ \\
\bottomrule
\end{tabular}}
\label{table:companion}
\end{table}

Now, looking at Table \ref{table:companion}, we notice that the $g-r$ and $r-i$ colours of the companion galaxy differ approximately 0.3 dex from the colours of the components of NGC 3115 (see Fig. \ref{fig:1Dcolor_gradient}), indicating that they probably do not have similar stellar populations. We could speculate that the companion galaxy suffered a recent star formation episode caused by the interaction with NGC 3115, given its blue colours. Unfortunately, this galaxy is too faint to be observed in the other bands or to perform SED fitting to get the SFR and age, so to confirm or discard this possibility. 

\begin{figure}
\hspace*{-0.5cm}
    \includegraphics[width=1.1\columnwidth]{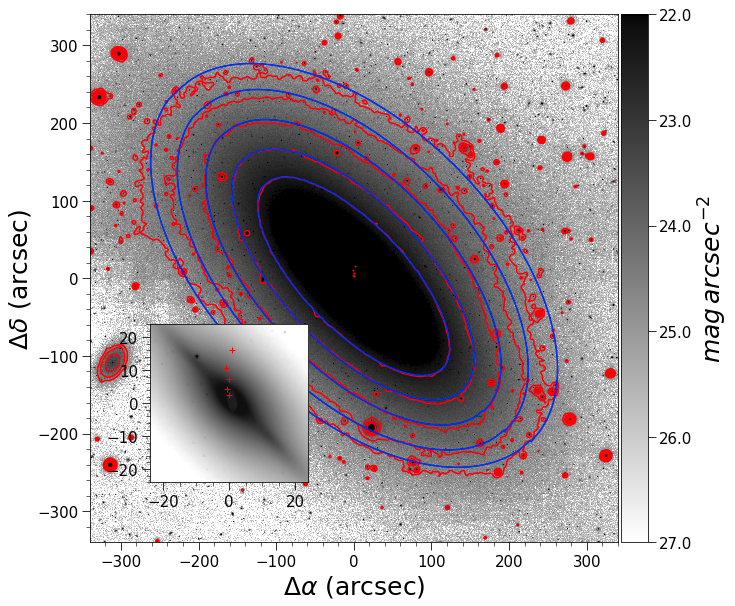}
    \hspace*{-0.5cm}
    \includegraphics[width=\columnwidth]{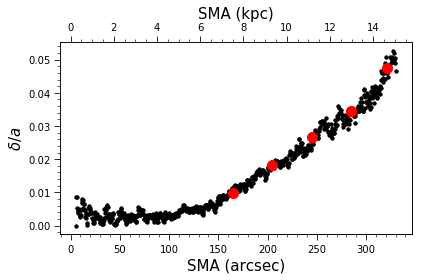}
    \caption{The off-centred core of NGC 3115. Top panel: i-band image of NGC 3115. Selected outer isophotes (in red) with their respective fitted ellipses (blue), the red plus symbols are their centres. The inset is a zoom-in of the inner part of the galaxy to clearly show the displacement of the ellipse centres. Bottom panel: The $\delta$/a ratio profile. The red points are the plotted isophotes. }
\label{fig:jose}
\end{figure}

Despite the need for deeper data to understand the stellar populations of both the companion KK084 and spiral arms and their relation to a possible encounter with NGC 3115, we can still probe the hypothesis of this encounter by analysing if there exists an offset in the core of NGC 3115, typical of recent interactions.
In fact, it has been shown that the external parts of bulges and elliptical galaxies can have a crossing time longer than the time of an interaction \citep{Combes1995, Binney2008}. Therefore, these regions suffer a tidal impulse in response to an encounter \citep{Aguilar1985, Binney2008} rather than a differential tidal one,  in which the internal dynamical time is shorter than the encounter time. Then, the external parts, due to an impulse answer, expand, while the core is displaced, as a tidal typical answer.
This offset can be up to $20\%$ of the observable radius of the galaxy \citep{Lauer1986, Lauer1988, Davoust1988, Combes1995, Gonzales-Serrano2000, Mora2019}.  Such offsets are excellent indicators of recent galactic encounters \citep{Combes1995, Mora2019}. To search for signs of recent interactions, we performed a photometric analysis following \cite{Mora2019}, who analysed the distortions of the Penguin system (Arp 142). We modelled the i-band image of NGC 3115 with the {\sc ELLIPSE} task in IRAF, letting all geometrical parameters, such as position angle, ellipticity and centre of the ellipses unconstrained. In order to quantify the displacement, we used $\delta/a$ ratio \citep{Lauer1988, Mora2019}, where $\delta$ is the separation of the isophotal centre to the nucleus and $a$ is the length of the semi-major axis (SMA). In Fig. \ref{fig:jose}, we show the radial profile of $\delta/a$ and the outermost isophotes of NGC 3115 and respective ellipses fitted, and their centres.  We can see, in the bottom panel of Fig. \ref{fig:jose}, that from about 160 arcsecs from the centre of the semi-major axis (SMA), the centre of the isophote starts to show an offset towards the South, revealing a maximum displacement of about 5\% ($\approx 800$ pc). Actually, the direction of the offset could serve as a constraint to the location of the pericentral passage, e.g. \citet{Combes1995, Mora2019}. This is a strong indication that NGC 3115  had a pericentre passage with a companion galaxy just a few hundred thousand years ago \citep{Combes1995, Mora2019}.
A natural candidate as the disturbing agent would be KK084. From the magnitudes measured by \cite{Sharina}, we have estimated a mass for KK084 of M$_{\star} = 1.5 \times 10^8$ M$_{\odot}$. Such a low mass would imply a high-speed encounter with a small impact parameter for
the interaction between NGC3115 and KK084 \citep{Aguilar1986}. Moreover, KK084 might have lost part of its mass during previous passages around NGC 3115 \citep{Aguilar1986,Mayer2001}.

\subsection{The Formation History of NGC 3115}
\label{sec:formation}

\begin{figure*}
    \centering
    \includegraphics[width=\textwidth]{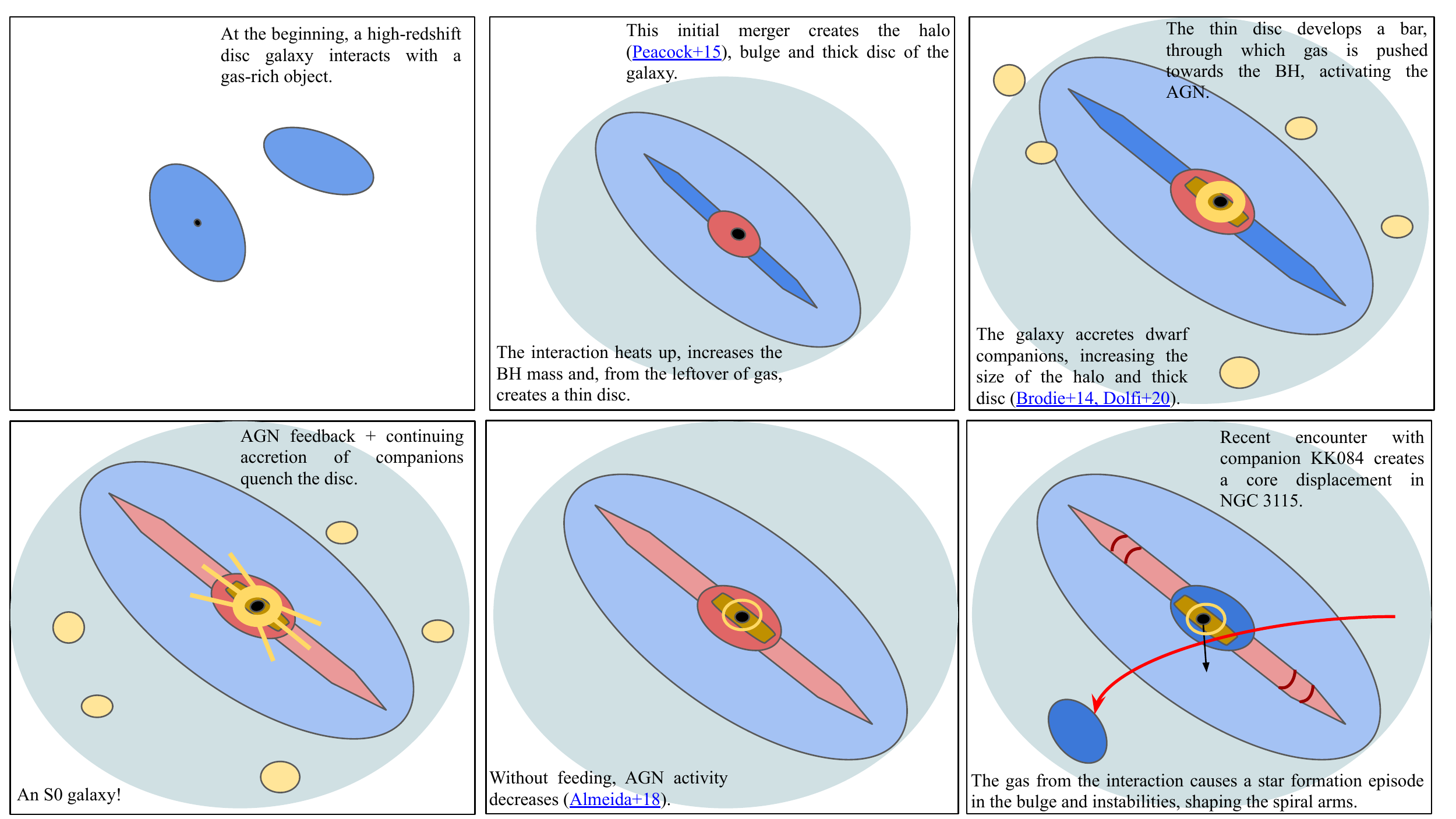}
    \caption{Formation scenario proposed for NGC 3115. The first panel shows two high-redshift galaxies, one with a central black hole. The second panel shows the moment after the encounter between the two galaxies, forming the bulge (red ellipse), thick disc (light blue ellipse) and halo (light green circle), and, afterwards, the formation of the thin disc (blue equatorial component). The third panel shows the formation of the bar (mustard equatorial rectangle), igniting the AGN activity (yellow central ring). Companion dwarf galaxies (yellow circles) are accreted, increasing the thick disc and the halo of the galaxy. The fourth panel shows the effect of AGN feedback (yellow lines coming out of the bulge) and the quenched thin disc (light pink equatorial component). The fifth panel shows the nearly passive galaxy with low AGN activity. The sixth panel shows the passage of the dwarf companion KK 084 (left blue ellipse), the direction of displacement of the core due to the passage (black arrow), its possible trajectory (red arrow), and the effect of the passage: bulge becoming bluer in its centre (central blue ellipse) and formation of spiral-like features (blue semi-circles in the middle of the thin disc).}
    \label{fig:formation}
\end{figure*}

NGC 3115 is a unique and extremely interesting object. Studying this galaxy, we found several extra components to the system, even when other lenticular galaxies usually present only a bulge, a disc and a bar \citep[or a lens][]{Johnston20}. The existence of these extra components in NGC 3115 (nuclear ring, end-on bar, spiral-like features) reinforce a complex formation scenario for this galaxy. Including data from an extended wavelength range in the analysis brings us important information regarding the stellar populations of each of these components.
As previously discussed, NGC 3115 was largely studied in the past, using photometric, spectroscopic, discrete tracers and IFU data.
\citet{Cortesib}, using PNe as tracers of the overall stellar kinematics, were able to rule out the scenario in which NGC 3115 is simply a stripped spiral galaxy, given the high value of $V/\sigma \simeq 3.3$ in the thick disc of the galaxy. The same result was obtained by studying the globular clusters kinematics \citep{Zanatta}. Reinforcing this point, \cite{Falcon-Barroso} show, using CALIFA data of 300 galaxies across the Hubble sequence, that spiral galaxies hardly fade into S0s, unless specific conditions of light concentration, angular momentum and $V/\sigma$ are satisfied (as for Sa galaxies).  In the other cases, pre-processing might have taken place in the formation of the lenticular, either through tidal interactions or mergers. 

Various works proposed the two-phase formation scenario as the most plausible to explain the formation of this galaxy \citep{Arnold11,Arnold14,Guerou,Dolfi}. 
\cite{Arnold11} described this formation scenario as the central region being formed in a violent and dissipative event, such as a gas-rich major merger, followed by a sequence of dry minor mergers and accretion events, that would have been responsible for the assembly of the outer spheroid and halo of the galaxy. Such a scenario is supported by the declining rotation and metallicity profiles measured for this galaxy, in their work, using optical imaging and multi-slit spectra. 
\cite{Guerou}, using IFU data, recovered the same kinematics and metallicity profiles as \cite{Arnold11} and \cite{Arnold14}. In their paper, \cite{Guerou} also hinted at the possibility of new components to this galaxy. They suggest the presence of an end-on bar,  a nuclear disc (suggested by the correlation between the mean and skewness of the measured line-of-sight velocity distribution), and of remnant spiral arms in the faint end of the thin disc of the galaxy.

\cite{Poci} suggest that the bulge and subsequently the thin disc of NGC 3115 could have been formed via a ``Compaction'' \citep{Dekel,Zolotov} scenario at early times, where a dissipative contraction would bring gas onto the central region of the galaxy via cold streams, which could trigger star formation events in its centre, being rapidly quenched in such massive galaxies as NGC 3115 \citep{Zolotov}. 
They also confirmed that the dynamics and stellar population of the thick disc is the result of the accretion of dwarf companions and satellites, which may increase the thick disc dispersion velocity and flatten the metallicity gradient.
In a more recent study, \cite{Dolfi}, using a large amount of data on NGC 3115 and faint companions, covering up to 6.5 effective radii and implementing an innovative kinematic method, confirmed the assembly history proposed by \cite{Arnold11}, finding an older pressure-supported bulge, followed by a transition in the kinematic profile at $0.2\, Re$ to a rotationally-supported disc, that goes out to 2-2.5 effective radii and then changes again to an outer pressure-supported spheroid. Such transitions agree well with the two-phase formation scenario.

Moreover, the stellar halo mass of NGC 3115 has been estimated to be nearly 14\% of the total stellar mass of the galaxy \citep[e.g.][]{Peacock}. This could be strong evidence that a merger had a key role in the formation of this galaxy. The formation of the thick disc could be attributed to this initial merger, although other mechanisms could also be responsible for that, e.g. clump instabilities in the early disc -- see \citet{Bournaud09,Clarke,BeraldoSilva}.

The results of this work, from the analysis of the SED fitting of the galaxy's components and the galaxy colour gradients, agree with the previously proposed scenario for the formation of NGC 3115. We see that all of the components of the galaxy are old, as it is retrieved by \cite{Guerou} and \cite{Dolfi}, and we also recover the declining metallicity profile from the bulge out to the thick disc (outer spheroid), as recovered by \cite{Arnold11} and \cite{Poci}. 

One difference, however, between our work and the previously cited works, is that the inclusion of images in the ultraviolet has suggested the presence of blue stellar populations in the bulge of the galaxy, that we propose to be the result of star-formation enhancement by an encounter with the companion KK084 that was capable, on one hand, to increase the star formation in the bulge and, on the other hand, to cause instabilities in the disc, generating spiral-like/ ring-like features in the faint end of the thin disc. This interaction would have caused a core displacement in NGC 3115, as discussed in Section \ref{sec:dwarf}.

The process of rejuvenating S0 galaxies has been largely studied in the past few years \citep{Mapelli,Bresolin, Delgado, Barway}. In fact, \cite{Mapelli} showed, using simulations, that recent encounters of S0 galaxies with gas-rich satellites usually form gas-rings and can reignite star formation in the central region of galaxies, rejuvenating the S0. They show that minor mergers may trigger episodes of star formation in the S0s that would last for $\sim 10$ Gyr, which could explain both the presence of the ring-like features and the recent star formation in the bulge of NGC 3115.

Incorporating these results into the previous scenarios proposed for its formation, we propose a novel scenario for the formation of NGC 3115: initially, a violent, dissipative gas-rich major merger created the bulge, halo and thick disc of the galaxy, then the remaining gas from the merger was able to generate the observed thin disc. Instabilities in the disc would have created the central bar, and such bar would be responsible for dragging gas from the disc to the centre of the galaxy, feeding the SMBH and igniting the AGN activity. With time, the galaxy starts accreting companions, which, in turn, enrich and increase the size of the now-massive thick disc, resulting in a disc with a mixed population of accreted stars and stars formed in-situ.
The continuous accretion and increasing mass of the galaxy cause feedback events, ceasing the star formation in the galaxy. The now gas-less thin disc is not able to feed the AGN any further, diminishing the AGN activity (observed to be low nowadays, \citealt{Almeida, Jones}) and the galaxy enters a phase of passivity. A recent encounter with the companion KK084 could have been capable of creating a starburst event in the centre of the galaxy, explaining the blue population and high SFR observed, and have also caused instabilities in the disc that would force the stars to rearrange in the spiral-like features observed today. Furthermore, the passage of the companion KK084 would have caused a core displacement to the south in NGC 3115 just a few hundred thousand years ago, and the direction of this displacement may set a constrain in the trajectory of the companion galaxy.
This formation scenario is summarised in Fig. \ref{fig:formation}.

\section{Conclusion}

In this work, we study the field lenticular galaxy closest to the Milky Way, NGC 3115.
For this purpose, we use 11 images from UV to IR wavelengths, observed with different telescopes, to perform a multi-wavelength broad-band fitting.
Using {\sc GALFITM}, we model this galaxy with a set of initial structural parameters to find the ones that best fit the galaxy and its components. The best-fit model of NGC 3115 has three components: a bulge with S\'ersic profile described by $n=3.5$ and $Re = 21.69$ arcsec ($0.97$ kpc), a thick disc and a thin disc, whose disc scale length increase linearly with wavelength. Given the high value of the axis ratio of the thick disc ($b/a = 0.44$), it might be interpreted as an oblate spheroid.
We also find evidence of the presence of a bar and spiral-like features from the residuals of these fits.
We find that the age of the galaxy as a whole and its components are around $\simeq 11$ Gyr, in agreement with previous works. 

In particular, we recover the same age/metallicity gradients as found by \cite{Guerou} and \cite{Poci} using IFU optical data. The results of this work are also consistent with the NUV-r results from \cite{Kaviraj} and the S\'ersic index ($n$) vs. BH mass relation found by \cite{Dullo} and \cite{Graham}. 

In the present analysis, by recovering the star formation history of the galaxy components, after performing SED fitting using {\sc CIGALE}, we show, for the first time, that the galaxy may have had a last burst at a few Myr ago, during a pericentre passage of the companion KK084, resulting also in a core displacement of the lenticular to the South, where the direction of this displacement can be used to constrain the trajectory of the companion around the lenticular.

We propose a formation scenario for NGC 3115 that incorporates results from previous works \citep{Guerou,Poci,Arnold11,Arnold14,Dolfi,Zanatta,Cortesib,Peacock} with new findings from this work, as explained at the end of the Section \ref{sec:formation} and summarised in Fig. \ref{fig:formation}.

We conclude that multi-band fitting of galaxy images followed by multi-component SED fitting are powerful tools to recover galaxies' formation histories and physical properties, consistent with spectroscopic studies. In this sense, the novel multi-band surveys S-PLUS, J-PLUS and J-PAS provide a new window to the understanding of galaxy evolution, especially when complemented with archive all-sky surveys (as GALEX and WISE).

\section{Acknowledgments}
M. L. B., A. C. and C. M. de O. acknowledge the financial support provided by the S\~ao Paulo Research Foundation (FAPESP) (grants 2018/09165-6 and 2019/2388-0).
A. C. acknowledge the financial support provided by CAPES.

J. A. H. J .thanks to Brazilian institution CNPq for financial support through postdoctoral fellowship (project 150237/2017-0) and Chilean
institution CONICYT, Programa de Astronom\'ia, Fondo ALMA-CONICYT 2017, C\'odigo de proyecto 31170038.

V.P.D. and L.B.S. are supported by STFC Consolidated grant \#ST/R000786/1.

A. J. R. was supported by the Research Corporation for Science Advancement as a Cottrell Scholar.

The simulations in this paper were run at the High-Performance Computing Facility of the University of Central Lancashire and at the DiRAC Shared Memory Processing system at the University of Cambridge, operated by the COSMOS Project at the Department of Applied Mathematics and Theoretical Physics on be- half of the STFC DiRAC HPC Facility (www.dirac.ac.uk). This equipment was funded by BIS National E-infrastructure capital grant ST/J005673/1, STFC capital grants ST/H008586/1 and STFC DiRAC Operations grant ST/K00333X/1. DiRAC is part of the National E-Infrastructure.

This research has made use of the NASA/IPAC Extragalactic Database (NED), which is operated by the Jet Propulsion Laboratory, California Institute of Technology, under contract with the National Aeronautics and Space Administration, and of {\sc Montage}, funded by the National Aeronautics and
Space Administrations Earth Science Technology Office,
Computation Technologies Project, under Cooperative
Agreement Number NCC5-626 between NASA and the California Institute of Technology. {\sc Montage} is maintained by the NASA/IPAC Infrared Science Archive.

\textit{Additional software}: Astropy \citep{2013A&A...558A..33A, 2018AJ....156..123A}, Matplotlib \citep{4160265}, NumPy \citep{5725236,numpy2}.

\section{Data availability}
The data used in this work is available via the GALEX, Subaru Suprime Cam, DECam, 2MASS and WISE mission archives.

\bibliographystyle{mnras}
\interlinepenalty=10000
\bibliography{paper.bib}

\appendix

\input{appendix}

\bsp	
\label{lastpage}
\end{document}

%% file: appendix.tex
\section{GALFITM models}
\label{sec:appendix_galfitm}

The fundamental steps in fitting a galaxy and correctly decompose its light into its main components using GALFITM are: identifying the number of physically motivated components, identifying the order of the series associated with each parameter for every component and defining the initial conditions of each parameter. The components of a galaxy are usually defined by some key parameters, such as S\'ersic index (n), effective radius (Re), disc scale-length (Rs), position angle (PA), axis ratio (b/a) and diskyness/skewness (C0).
Each of these parameters in GALFITM is expanded in a wavelength-dependent series, and the variation of these parameters with the wavelength is defined by the order of this series (orders). As previously explained, this order can vary from the most restrict case, i.e. orders = 0, where all bands are tied to the initial condition, to the freest case, i.e. orders = number of fitted images (11, in our case), where this parameter can vary freely between bands. The number of fitted parameters will increase with the number of components, and so determining the best combination of the order of the series of each parameter can be a hard task, such as manually combining these orders and evaluating each output can take a lot of time.
In order to find this best combination, we created several GALFITM models using different combinations of orders of these parameters. This was done for models with one, two and three components.

For the models with one component, since we are not decomposing the galaxy in subcomponents, and, therefore, not separating its different colours, one might expect that the series would need to vary with a larger order, to account for differences in the structural parameters of the galaxy. Therefore, for the models with one component, we let five parameters vary with the order spanning from 0 to 7, creating models with each combination of these orders. These parameters were n, Re, PA, b/a and C0. This led to a total of 32768 models.

For the models with two components, we were more restrictive, since we are now decomposing the galaxy in at least two colours, we expect that the variation would be smaller between bands, and, therefore, the orders were only allowed to span from 0 to 3. Also, we can assume that there is no significant change among bands in the position angle or axis ratio of these components, therefore, we fix the order of these parameters to be 1, i.e. constant between bands, for both components. Moreover, analysing the models with one component, we realised that the C0 parameter was not affecting our results, so we excluded this parameter from the further modelling. We now have three parameters: n, Re and Rs, and their order can vary from 0 to 3, leading to 64 models.

Finally, for the model with three components, applying the same cuts as we did for the model with two components, we have four parameters that we allow to vary their order spanning from 0 to 3, these are n, Re, Rs-disc1 (hereafter Rs1) and Rs-disc2 (hereafter Rs2), which led to 256 models.

The models created and the orders of the series that each parameter can vary with are summarised in Table \ref{tab:models_simulation}.

\begin{table}
    \caption{Order of the series that each parameter can vary to find the best fit model with GALFITM}
    \begin{threeparttable}
        \scalebox{0.95}{%
    \begin{tabular}{ccc}
    \toprule
      \textbf{Model}   & \textbf{Parameter} & \textbf{Orders} \\
      \midrule
      \midrule
       \multirow{5}{*}{SS} & n & 0,1,2,3,4,5,6,7 \\
                           & Re & 0,1,2,3,4,5,6,7\\
                           & PA & 0,1,2,3,4,5,6,7\\
                           & b/a & 0,1,2,3,4,5,6,7\\
                           & C0 & 0,1,2,3,4,5,6,7\\ 
        \# of combinations - 1 comp & \multicolumn{2}{c}{32768}  \\ \hline    
       \multirow{5}{*}{BD} & n & 0,1,2,3 \\
                           & Re & 0,1,2,3 \\
                           & PA $_{\rm b \& d}$* & 1 \\
                           & b/a $_{\rm b \& d}$ & 1 \\
                           & Rs & 0,1,2,3 \\ 
        \# of combinations - 2 comps & \multicolumn{2}{c}{64}  \\ \hline
       \multirow{6}{*}{BDD} & n & 0,1,2,3 \\
                            & Re & 0,1,2,3 \\
                            & PA $_{\rm b \& d}$ & 1 \\
                            & b/a $_{\rm b \& d}$ & 1 \\
                            & Rs1 & 0,1,2,3 \\ 
                            & Rs2 & 0,1,2,3 \\ 
        \# of combinations - 3 comps & \multicolumn{2}{c}{256}  \\ 
        \bottomrule
    \end{tabular}}
    \begin{tablenotes}
     \small
     \item * the subscript {b\&d} denotes that these orders are used both for the bulge and disc(s) components.
    \end{tablenotes}
    \end{threeparttable}
    \label{tab:models_simulation}
\end{table}

These models were firstly analysed taking into consideration three statistics that we find representative of the goodness of the fit, these are:
\begin{itemize}
    \item Model/Input: this value is obtained by dividing the model images by the input images and taking the median of the resulting image. The concept behind this is to understand how well the model is mimicking the input image. Consequently, the best models are those who have Model/Input closest to one.
    
    \item Residual/Input: this is obtained by dividing the residual image by the input image in each band and taking the median of the result. With these statistics, we can understand how significant the residual fluxes are with respect to the data. The best models are expected to have values of Residual/Input close to zero, where, if this is smaller than zero, means that the galaxy is being over-fitted, and if higher, under-fitted.
    
    \item Residual/Sigma: these final statistics are obtained by dividing the residual image by the error of the input data. The idea is to understand how significant the residuals are concerning the error. Therefore, if sigma $>>$ residuals, the residuals are well within the error and this is a good fit. If residual $>$ sigma, the residuals are bigger than the errors and this is not considered such a good fit. Therefore, the best models will have the smallest Residual/Sigma.
\end{itemize}

\begin{figure*}
    \centering
    \includegraphics[width=0.9\textwidth]{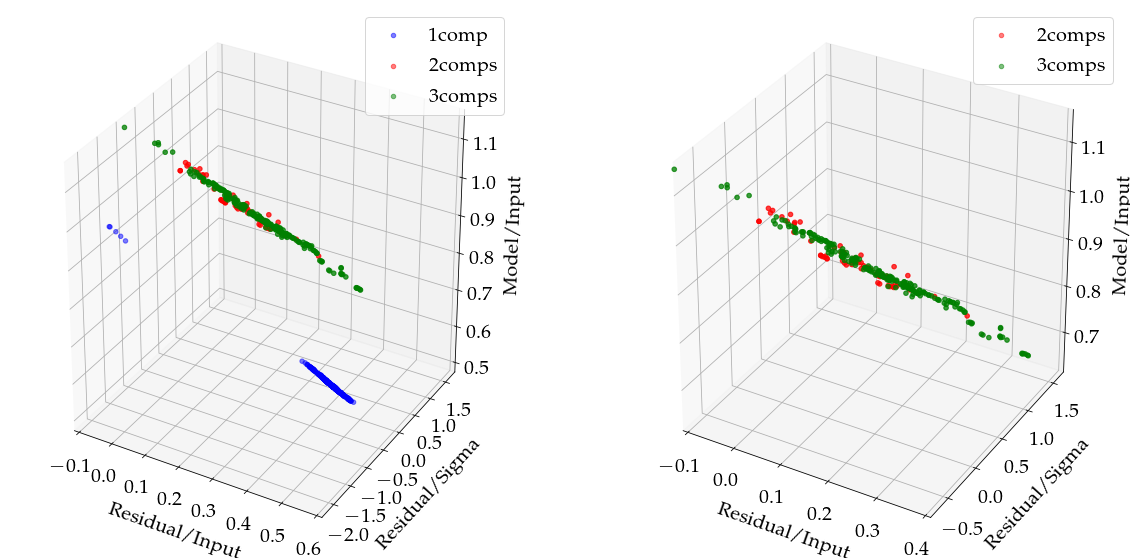}\\[0.5cm]
    \includegraphics[width=0.9\textwidth]{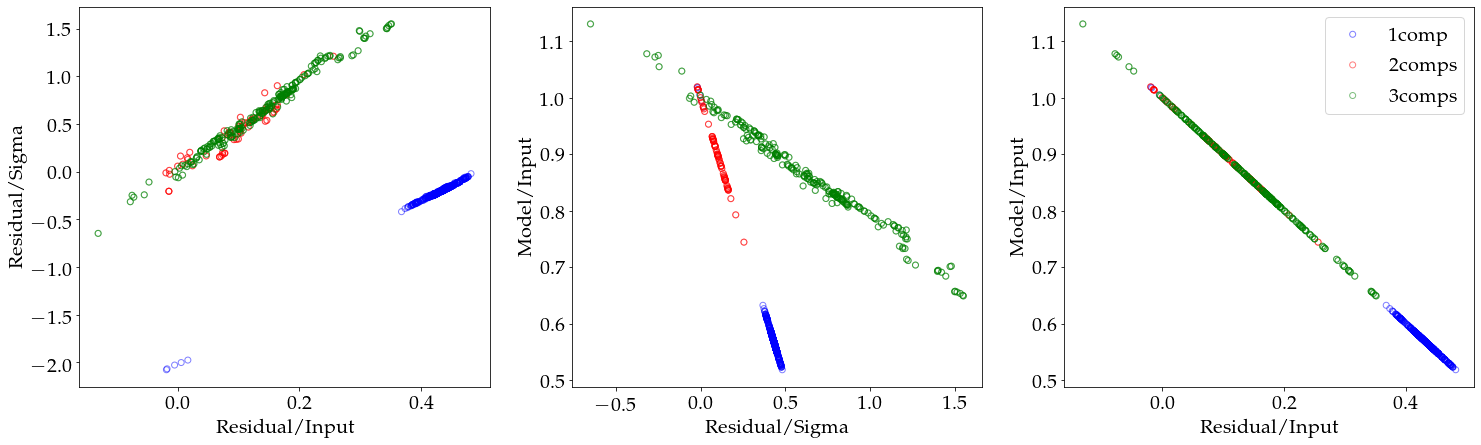}
    \caption{GALFITM models created for NGC 3115 using one, two and three components, with different combinations of orders that define how each parameter can vary with the wavelength. Models are evaluated according to the statistics: Residual/Input (how the residual compares with the input image: the smallest the better), Residual/Sigma (how the residual compares to the error of the input image, the smallest the better) and Model/Input (how well the model is mimicking the input image: the closest to one the better). The bottom line shows the 2D projection of each pair of axes of the 3D plot for the models with one, two and three components.}
    \label{fig:3D_comps}
\end{figure*}

We also analyse our models in terms of the Bayesian Information Criteria and Akaike information criteria. However, these methods are taken less into account, since they are strongly biased by the best $\chi^2$, which not necessary resembles the best fit model, since we are looking for models with physically meaningful components, and sometimes the models with the lowest $\chi^2$ are found by creating mathematical artefacts. 
In Fig. \ref{fig:3D_comps}, we show the distribution of the models with one, two and three components in the 3D plane, shaped by the Model/Input, Residual/Input and Residual/Sigma statistics. 

\begin{figure*}
    \centering
    \includegraphics[width=0.8\textwidth]{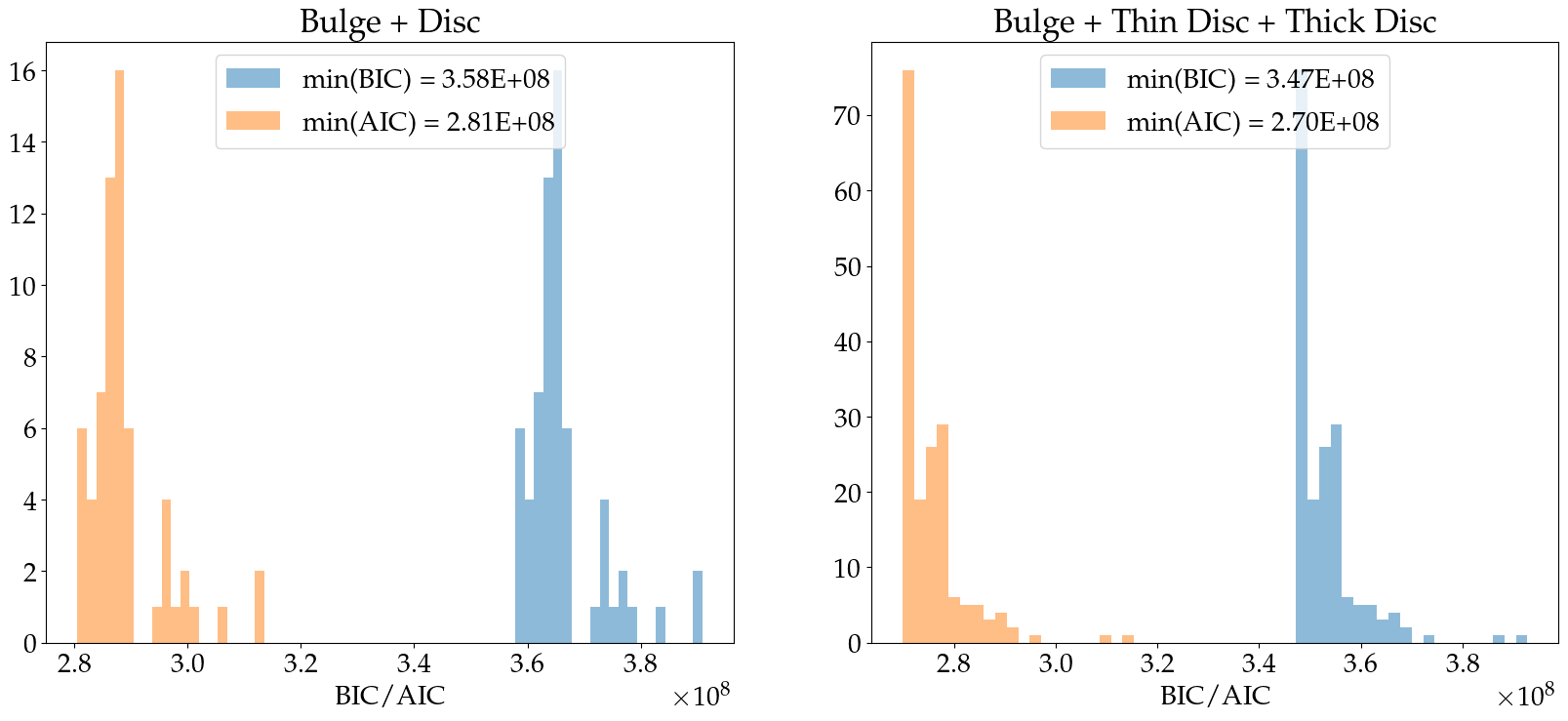}
    \caption{BIC and AIC of the models with two and three components, where the different models are created by using different combinations of orders of the wavelength-dependent series that define the variation of the parameters among bands.}
    \label{fig:hist_components}
\end{figure*}

From the left-hand side of Fig. \ref{fig:3D_comps}, we can directly understand that models with one component are not sufficient to recover the total light of NGC 3115, as they fall substantially lower than the models with two and three components in the Model/Input axis, meaning that the model found are comprising about half of the light of the input images.
In the right-hand side of Fig. \ref{fig:3D_comps}, we focus on the models with two and three components. As we can see, the majority of these models share a locus in the plot. Both models with two and three components behave quite well in the three statistics. We can see, however, that there are about 5 models with three components that outstand this locus, presenting values close -or exactly the expected- for the three statistics, i.e. Model/Input $\approx 1$, Residual/Input $\approx 0$ and Residual/Sigma $\approx 0$.
To further investigate the difference between the models with two and three components, we analyse in Fig. \ref{fig:hist_components}, the BIC and AIC of these models.
Confirming what we found previously, what we can see in Fig. \ref{fig:hist_components} is that the models with three components (Bulge + Thin Disc + Thick Disc) deliver the minimal value of BIC and AIC, representing the best fits for the data.

\begin{figure*}
    \centering
    \includegraphics[width=0.85\textwidth]{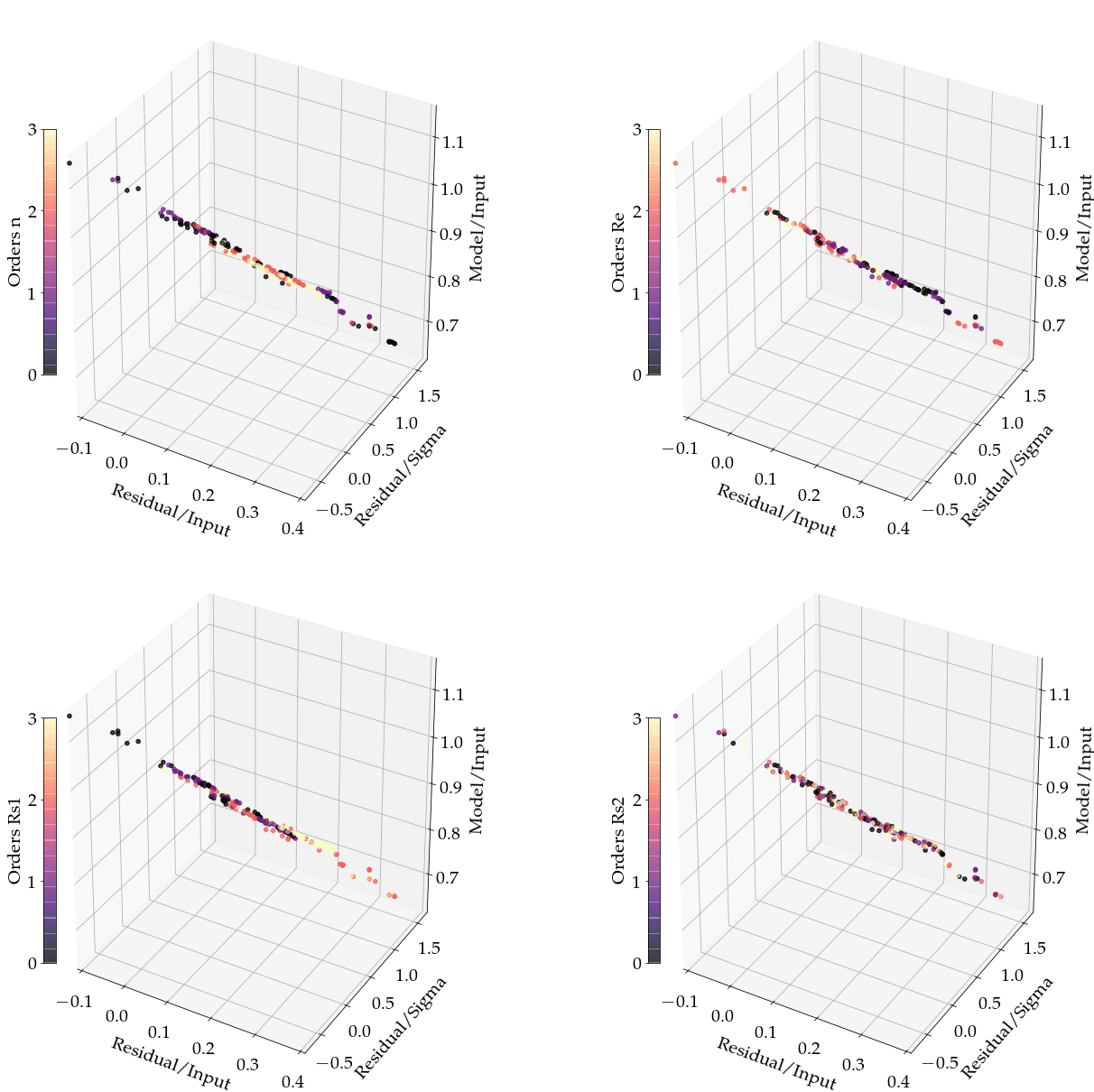}
    \\[0.5cm]
    \includegraphics[width=0.85\textwidth]{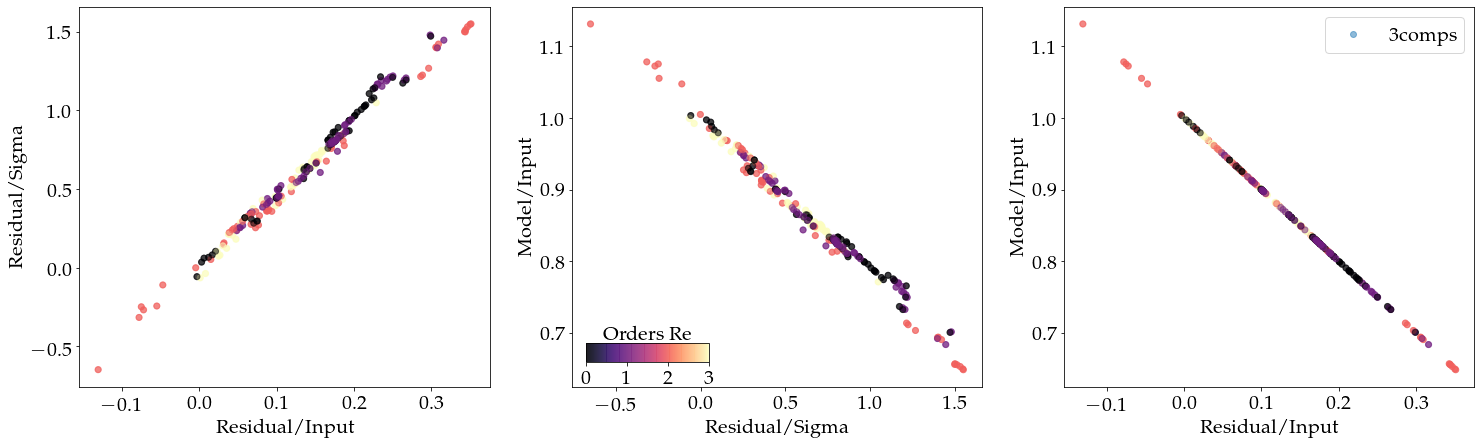}
    \caption{GALFITM models created for NGC 3115 using three components and different combinations of orders of the parameters. Models are evaluated according to the statistics: Residual/Input (how the residual compares with the input image: the closest to zero the better), Residual/Sigma (how the residual compares to the error of the input image, the closest to zero the better) and Model/Input (how well the model is mimicking the input image: the closest to one the better). \textit{Upper left:} distribution of models coloured with the orders that define the variation of the Sérsic index among bands. \textit{Upper right:} distribution of models coloured with the orders that define the variation of the effective radius among bands. \textit{Middle left:} distribution of models coloured with the orders that define the variation of the disc scale-length of the first disc among bands. \textit{Middle right:} distribution of models coloured with the orders that define the variation of the disc scale-length of the second disc among bands. \textit{Bottom line:} the 2D projection of each pair of axes of the 3D plot. These panels are coloured with the orders of variation of the effective radius, such as the top left panel.}
    \label{fig:3D_params}
\end{figure*}

Now, to better understand the variation of these three component models in the plane, we show in Fig. \ref{fig:3D_params}, the variation of the models according to the order of the series of each parameter (n, Re, Rs1 and Rs2).

There are insightful conclusions that we can take from Fig. \ref{fig:3D_params}. Firstly, analysing the first panel, we can see that the 5 best-fit models are the ones where the S\'ersic index can either be fixed to the initial condition (orders = 0) or constant among bands (orders = 1), showing that, when we carefully separate the galaxy into components, the parameters that define each component do not vary significantly between bands, because we are most likely not dealing with mixed colours/structural parameters anymore. The radius, however, as seen from the other panels in Fig. \ref{fig:3D_params}, can still have some small variation (i.e. orders =1, constant or orders =2, linear) among bands, as it would be expected, since we know from observing the galaxy, that its radius severely changes from UV to IR. Most importantly, from all panels, what we see is that large freedom of the parameters seems to worsen the models, reinforcing once again the power of GALFITM to tie the images together and deliver reliable and consistent physical parameters.

From this analysis, we can take several important information that the best model of NGC 3115 must comprise:
\begin{itemize}
    \item Morphology: three components(a bulge and two exponential discs).
    \item Orders of the parameters: n, PA and b/a should either be fixed to the input value or constant among bands. Re, Rs1 and Rs2 can either be constant or vary linearly, but not more than that.
    \item The parameters recovered in the best fit model in this simulation can be used as reliable initial conditions to future models. 
\end{itemize}

Although the best fit model found by these simulations is not the one chosen as the best for this work, with this analysis we recovered the three fundamental steps, as described in the first paragraph of this appendix, to create the most consistent and robust GALFTIM model, described in the main body of the paper and used for our analysis.